\documentclass[aps,prl,twocolumn,superscriptaddress,floatfix]{revtex4-2}
\usepackage{graphicx, bm, amsmath, amssymb, color}
\usepackage{amstext, amsfonts}
\usepackage{float}
\usepackage{algpseudocode}
\usepackage{multirow, physics, tasks}
\definecolor{darkblue}{rgb}{0.184, 0.192, 0.584}
\usepackage[colorlinks=true,linkcolor=blue,
urlcolor=black,anchorcolor=darkblue,
citecolor=blue]{hyperref}
\usepackage[normalem]{ulem}
\usepackage[dvipsnames]{xcolor}
\usepackage[linesnumbered,ruled,vlined]{algorithm2e}
\usepackage{array}
\usepackage{booktabs}
\usepackage{appendix}
\usepackage{cleveref}
\usepackage{xr}
\graphicspath{{figures/}}

\def\BibTeX{{\rm B\kern-.05em{\sc i\kern-.025em b}\kern-.08em
    T\kern-.1667em\lower.7ex\hbox{E}\kern-.125emX}}

\begin{document}

\title{Enhanced Quantum Signal Control and Sensing  Under Multicolored Noise via Generalized Filter Function Framework}
\author{Zhi-Da Zhang}
\author{Yao Song}
\affiliation{Shenzhen International Quantum Academy (SIQA), Futian District, Shenzhen, P. R. China}
\affiliation{Shenzhen Institute for Quantum Science and Engineering (SIQSE),
Southern University of Science and Technology, Shenzhen, P. R. China}
\author{Wen-Zheng Dong}
\affiliation{Shenzhen International Quantum Academy (SIQA), Futian District, Shenzhen, P. R. China}
\author{Xiu-Hao Deng}
\email{dengxh@sustech.edu.cn}
\affiliation{Shenzhen International Quantum Academy (SIQA), Futian District, Shenzhen, P. R. China}
\affiliation{Shenzhen Institute for Quantum Science and Engineering (SIQSE),
Southern University of Science and Technology, Shenzhen, P. R. China}
\begin{abstract}
We introduce a generalized filter-function framework that treats noise coupling strength as a tunable control parameter, enabling target noise suppression across user-defined frequency bands. By optimizing this generalized filter function, we design band-selective control pulses that achieve $0.9999$ fidelity of single- and two-qubit gates under strong noise with diverse spectral profiles. We further extend the method to selectively enhance the signal-to-noise ratio for quantum sensing of AC signals with an enhanced precision of up to $10$ dB. The resulting control pulses are experimentally feasible, offering a practical pathway toward robust quantum operations and high-precision signal processing under spectrally complex noises.
\end{abstract}
\maketitle

In quantum technologies, noise often exhibits diverse spectral characteristics, which severely constrain high-fidelity gate operations and precise sensing~\cite{cheng2023noisy}. For instance, low-frequency $1/f$ noise dominates in many solid-state qubit platforms, while high-frequency noise arises from quantum environmental interactions or high-energy excitations~\cite{wellard2002thermal,pedrocchi2015majorana,alicki2002dynamical,yoneda2018quantum,paladino20141,kumar2016origin,brownnutt2015ion}. Techniques such as dynamical decoupling (DD)~\cite{meiboom1958modified,khodjasteh2007performance}, composite pulses (CP)~\cite{merrill2014progress,brown2004arbitrarily}, and dynamically corrected gates (DCG)~\cite{khodjasteh2009dynamically,hai2022universal,yi2024robust} have been developed to suppress quasi-static noise. To address time-dependent noise, the filter function formalism provides a framework for generalizing the robust control protocols~\cite{wang2014robust,wang2014noise}, enabling further suppression of correlated noise~\cite{yang2016noise,uhrig2007keeping,martinis2003decoherence,kofman2001universal,kabytayev2014robustness,cywinski2008enhance,ball2016effect}. Combining with the Magnus~\cite{magnus1954exponential,blanes2009magnus} or Dyson~\cite{dyson1949radiation} expansion, one can analyze quantum operation fidelity subject to colored noise by evaluating the convolution between the noise spectrum and the filter function~\cite{green2013arbitrary,green2012high,paz2014general,hangleiter2021filter,clausen2010bath,ball2021software,cerfontaine2021filter}. This approach has enabled both the optimization of robust pulse sequences for noise suppression~\cite{uys2009optimized,uhrig2007keeping,malinowski2017notch,dong2023resource,biercuk2009optimized,soare2014experimental,ball2015walsh,hangleiter2021filter,le2022analytic} and quantum noise spectral estimation~\cite{paz2017multiqubit,malinowski2017spectrum,frey2017application,dong2023resource,bylander2011noise,alvarez2011measuring}.

Despite these advances, most filter-function methodologies assume that the noise coupling remains fixed, thereby overlooking an important avenue for noise mitigation. In many hardware platforms, e.g. transmon qubits~\cite{cheng2023noisy} and fluxonium~\cite{manucharyan2009fluxonium}, the system–noise coupling strength depends on the biased point and hence is controllable, offering additional leverage for frequency-specific noise suppression. We developed a generalized filter function framework that incorporates tunable coupling strength and employs an algorithm for complex and optional constraints optimization with autodifferentiation (COCOA) ~\cite{song2022optimizing}, to identify smooth, optimal control waveforms that adhere to realistic experimental constraints. We further define a noise susceptibility metric to quantify the efficacy of these optimized pulses against time-dependent noise. Our results reveal that, when coupling strength is tunable, noise filtering can be achieved with fewer control resources while preserving high gate fidelities and robust sensing with high signal-to-noise ratios. The proposed protocols are designed for experimental feasibility, requiring only moderate modifications to existing hardware controls. As such, this work provides a practical pathway toward noise-resilient quantum operations and improved signal processing across a wide spectrum of noise environments.

\paragraph{General filter function formalism.-}We begin with the filter function formalism for semiclassical treatment of noise sources, which can be extended to quantum noise through the comprehensive framework developed by Paz-Silva \textit{et al.}\cite{paz2014general}. Consider an open multi-qubit system with the Hamiltonian: $H(t)=H_s(t)+H_{ctrl}(t)+H_n(t) \equiv H_0(t)+H_n(t)$, where $H_s,H_{ctrl},H_n$ represent intrinsic system Hamiltonian, control Hamiltonian, and coupling-tunable noise Hamiltonian, respectively. In the operator basis $ {\Lambda_u} $ typically chosen to be tensor products of Pauli operators, the noiseless term $H_0(t)=\sum_u h_u(t)\Lambda_u$ and the noise term $H_n(t)=\sum_u\delta_u(t)\Lambda_u$. In the interaction picture defined by $ H_0 $, the noise generates the \textit{error evolution operator}\cite{hai2022universal} taking the form $U_e(t)=U_0(t)^\dagger(t) U(t) =\mathcal{T}{+}e^{-i\int{0}^{t}ds\widetilde{H}n(s)}$, where the noiseless evolution is $U_0(t)=\mathcal{T}{+}e^{-i\int_{0}^{t}dsH_{0}(s)}$ and the total evolution is $U(t)$. The noise Hamiltonian $ \widetilde{H}n(t) = U_0(t)^\dagger H_n(t) U_0(t) $ can be decomposed in a generalized form $\widetilde{H}n(t) \equiv \sum{u, v} \delta_u(t) , \Lambda{v} , \mathrm{Tr}{S}[U{0}^{\dagger}(t) \Lambda_{u} U_{0}(t) \Lambda_{v}] / d $ where $ d = 2^n $ is the system dimension.

The effective noise strength $\delta_u(t)$ encountered by the quantum system is associated with its physical origin $ \beta_j(t) $ through what we call the \textbf{coupling strength} $ c^{(j)}_u(t) $, as expressed by $\delta_u(t) = \sum_jc^{(j)}_u(t) \beta_j(t)$, with $j$ indexing a noise source. Here, we refer to the set of system operators ${\Lambda_u }_j$ that a single noise source $\beta_j(t)$ affects as the \textbf{coupling operators} for that noise source, representing the different interaction channels between the noise and the system. Meanwhile, the parameter $c^{(j)}_u(t)$ quantifies the coupling strength—how strongly the noise $\beta_j(t)$ influences a particular operator $\Lambda_u$. This coupling strength is what we can tune to control the noise dynamics. For example, $ \beta_j(t) $ could represent environmental charge or flux noise, while $ \delta_u(t) $ represents effective noise amplitude.

Through the definition above we can change the noise Hamiltonian from the sum of different system operators to the sum of different noise sources.  Then the interaction Hamiltonian takes the form:
\begin{equation}
\begin{aligned}
&\widetilde{H}_n(t)\equiv \sum_{j,v}y_{j,v}(t)\beta_j(t)\Lambda_{v},~~~\text{where}\\
&y_{j,v}(t)=\sum_{u}c_{u}^{(j)}(t)\text{Tr}_S[U^{\dagger}_0(t)\Lambda_{u}U_0(t)\Lambda_v]/d.
\end{aligned}
\end{equation}
Compared to $\delta_u(t)$ before, we have noise modulation by tuning $c(t)$. However, the selection of the appropriate expansion method depends on several factors, including the number of noise sources, whether the noise is present, and whether the coupling strength is controllable. Both expansions can be applied simultaneously when the system operators involved do not conflict. This is because the two expansions only affect the composition of the control function $y(t)$, which determines the filter function later, rather than the form of the interaction Hamiltonian $\widetilde{H}_n(t)$. 

The \textit{error evolution operator} can also take the Magnus expansion form, $U_e = e^{\sum \Omega_\alpha}$ , where the $\alpha$th order Magnus term $\Omega_\alpha$ is a time-ordered integral of $\widetilde{H}_n(t)$, over the volume $V_{\alpha}\equiv\{0\leq t_{\alpha}\leq t_{\alpha-1}\leq\ldots\leq t_{2}\leq t_{1}\leq T\}$
\begin{equation}
\Omega_{\alpha}=\sum_{\vec{j},\vec{v}}\int_{V_{\alpha}}d\vec{t} ^{\alpha}\sum_{\pi\in\Pi_\alpha}f^{[\alpha]}(\pi)\  y^{[\alpha]}_{\vec{j},\vec{v}}(\pi) \beta^{[\alpha]}_{\vec{j}}(\pi) \Lambda^{[\alpha]}_{\vec{v}}.
\end{equation}
$d\vec{t} ^{\alpha}= dt_1...dt_\alpha$, $y^{[\alpha]}_{\vec{j},\vec{v}} (\pi)\equiv y_{j_1v_1}(\pi(t_1))\cdotp\cdotp\cdotp y_{j_\alpha v_\alpha}(\pi(t_\alpha))$, $\beta^{[\alpha]}_{\vec{j}}(\pi) \equiv \beta_{j_1}(\pi(t_1))\cdotp\cdotp\cdotp \beta_{j_\alpha }(\pi(t_\alpha))$, $\Lambda^{[\alpha]}_{\vec{v}}=\Lambda_{v_1}\cdotp\cdotp\cdotp \Lambda_{v_\alpha}$. Where $\pi $ is a permutation within a set $\Pi_\alpha$ containing all the $\alpha$-permutations, and $f^{[\alpha]}(\pi)$ are functions representing the coefficient of the $\alpha$th order Magnus expansion, which contains the imaginary number $i$,  and signs for different permutations.

The average gate fidelity~\cite{nielsen2002simple} is defined by $\mathcal{F}_{avg}=\langle|\frac1d\ \mathrm{Tr}[U_G(T)^\dagger U(T)]|^2\rangle=\frac1{d^2}\langle|\mathrm{Tr}[U_e(T)]|^2\rangle$, where $U_G$ is the target gate, and $\langle\cdot\rangle$ represents the average over semi-classical noise. Performing a Taylor expansion to $U_e$, we obtain:
\begin{equation}\label{whole_fid}
\begin{aligned}
\mathcal{F}_{avg}&= \sum_{r,r'}\frac{1}{r!r'!} \cdot \frac1{d^2} \mathrm{Tr}[(\sum \Omega_\alpha)^r] \mathrm{Tr}[(\sum \Omega^\dagger_\alpha)^{r'}]\\
&= \sum_{}\frac{1}{r!r'!(2\pi)^{||\vec{\alpha}_r||_1+||\vec{\alpha}'_{r'}||_1}} \mathcal{O}^{[*]}.
\end{aligned}
\end{equation}
The final expression involves multiple summations on $r,r';\ \vec{\alpha}_r,\vec{\alpha}_r';\ \vec{j}^{\alpha_1}\cdotp\cdotp\cdotp\vec{j}^{\alpha_r},\vec{v} ^{\alpha_1}\cdotp\cdotp\cdotp\vec{v}^{\alpha_r},\vec{j}^{\alpha'_1}\cdotp\cdotp\cdotp\vec{j}^{\alpha'_{r'}},\vec{v} ^{\alpha'_1}\cdotp\cdotp\cdotp\vec{v}^{\alpha'_{r'}}$. The dimension of $\vec{\alpha}_r$ is decided by $r$, and the dimension of $\vec{j}^{\alpha_j},\vec{v}^{\alpha_j}$ is determined by the corresponding $\alpha_j$. The $2\pi$ term comes from Fourier transformation. We define the symbol $[*]$ covering all variables above, and the symbols $\mathcal{J, V}$ containing all vectors of $\vec{j},\vec{v}$ respectively. Here, $\mathcal{O}^{[*]}$ is:
\begin{equation}
\begin{aligned} 
\mathcal{O}^{[*]}
=\int & D\vec{\omega}\ G_{\mathcal{J, V}}(\vec{\omega}^{\alpha_1}\cdotp\cdotp\cdotp \vec{\omega}^{\alpha_r}\vec{\omega}^{\alpha'_1}\cdotp\cdotp\cdotp \vec{\omega}^{\alpha'_{r'}},T)\\
 &S_{\mathcal{J}}(\vec{\omega}^{\alpha_1},\cdotp\cdotp\cdotp ,\vec{\omega}^{\alpha_r},\vec{\omega}^{\alpha'_1},\cdotp\cdotp\cdotp ,\vec{\omega}^{\alpha'_{r'}})\\
 &\frac{1}{d^2}\mathrm{Tr}(\Lambda^{[\alpha_1]}\cdotp\cdotp\cdotp \Lambda^{[\alpha_r]})\mathrm{Tr}(\Lambda^{[\alpha'_1]}\cdotp\cdotp\cdotp \Lambda^{[\alpha'_{r'}]})
\end{aligned}
\end{equation}
$G_{\mathcal{J, V}}(\vec{\omega}^{\alpha_1} \cdots  \vec{\omega}^{\alpha'_{r'}}, T)=G^{[\alpha_1]}\cdots G^{[\alpha'_{r'}]}$  represents the general filter function for the noise spectral density $S_{\mathcal{J}}(\vec{\omega}^{\alpha_1}, \vec{\omega}^{\alpha_2}, \dots) = \int D\vec{t} \langle \beta^{[\alpha_1]}_{\vec{j}^{\alpha_1}}(\vec{t}^{\alpha_1}) \beta^{[\alpha_2]}_{\vec{j}^{\alpha_2}}(\vec{t}^{\alpha_2}) \dots \rangle e^{-i(\vec{\omega}^{\alpha_1} \cdot \vec{t}^{\alpha_1} + \vec{\omega}^{\alpha_2} \cdot \vec{t}^{\alpha_2} + \dots)}$. Here \(G^{[\alpha]} = G^{[\alpha]}_{\vec{j},\vec{v}}(\vec{\omega}^\alpha, T) \) is the \(\alpha\)-th order subfilter function:
\[
G^{[\alpha]}_{\vec{j},\vec{v}}(\vec{\omega}^\alpha, T) = \int_{V_\alpha} d\vec{t}^\alpha \sum_{\pi \in \Pi_\alpha} f^{[\alpha]}(\pi) \, y^{[\alpha]}_{\vec{j},\vec{v}}(\pi)e^{i\vec{\omega}^\alpha \cdot \pi(\vec{t}^\alpha)}.
\]
The term \( \frac{1}{d^2} \mathrm{Tr}(\cdots) \) acts as a weight factor, which can be 0 or 1.

For small noise amplitudes, we truncate Eq.~\eqref{whole_fid} to the second order. And assuming uncorrelated noise from different sources (\( S_{j_1,j_2}(\omega_1,\omega_2) = 0 \) for \( j_1 \neq j_2 \)) and Gaussian stationary noise (\( S_{j,j}(\omega_1, \omega_2) = 2\pi S_j(\omega_1)\delta(\omega_1 + \omega_2) \)), the fidelity simplifies to 
\begin{equation}
    \begin{aligned}
        \mathcal{F}_{\text{avg}} \approx 1 - \sum_j \int \frac{d\omega}{2\pi} F_j(\omega, T) S_j(\omega),
    \end{aligned}
    \label{approx_fid}
\end{equation} 
where \( F_j(\omega, T) = \sum_v |G^{[1]}_{j,v}(\omega, T)|^2 \) is the second order filter function for noise $\beta_j$. Suppressing \( F_j(\omega, T) \) leads to robust quantum gates, while amplifying specific frequency ranges enables high signal-to-noise ratio sensing.

Finally, we define the noise susceptibility. Assuming noise amplitude \( A \) such that \( S(\omega) \propto A^2 \), the overlap between \( F_j(\omega, T) \) and \( S_j(\omega) \) becomes: $\int \frac{d\omega}{2\pi} F_j(\omega, T) S_j(\omega) = C_j A_j^2$, where \( C_j \) is a constant determined by the filter function. If we want to quantify the susceptibility to one noise source, the fidelity approximation becomes $\mathcal{F}_{\text{avg}} \approx 1 -  C_j A_j^2$, and the logarithmic dependence satisfies $\log(1 - \mathcal{F}_{\text{avg}}) \approx 2\log(A_j) + \log\left(C_j\right)$, where the intercept of the linear relation gives \(\log( C_j)\). We define the \textbf{noise susceptibility} to noise source $j$ as $C_j$. This allows us to quantify waveform robustness under noise by comparing \( C_j \) for different noise spectra. These quantities serve as the cost function in the COCOA algorithm~\cite{song2022optimizing,supp} to optimize pulses for various tasks illustrated as follows.

\paragraph{Robust single-qubit gate.-}
We here focus on designing and implementing a noise-resilient $X_\pi$ gate in the presence of multi-band colored noise. Configuring the control field as a multi-band-stop filter enables the suppression of specific noise frequencies within one pulse. This approach demonstrates the flexibility of filter-function engineering for single-qubit gate robustness.

\begin{figure}[t]
  \centering
  \includegraphics[width=1\linewidth]{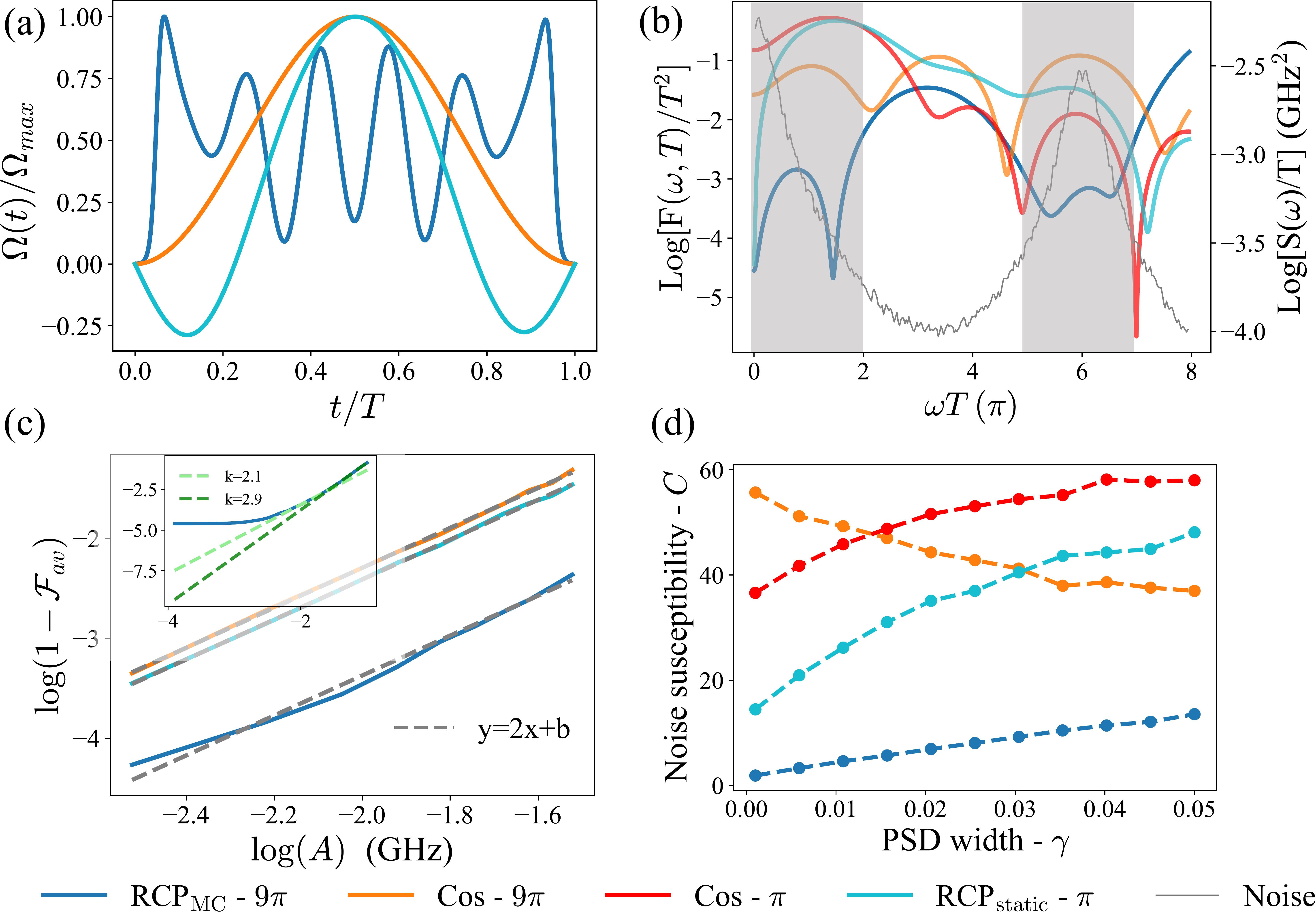}
  \caption{Comparisons of $X_\pi$ gates in the presence of multi-colored noise. (a) Different waveforms of the control field. (b) The corresponding filter functions of these waveforms, plotted alongside the noise spectrum (gray curve) and target bands (gray area). (c) log view of average X-gate fidelity implemented by mixed noise RCP, Cos $9\pi$-pulse, and static noise RCP against time-dependent noise with different amplitude.(The curve of Cos $\pi$-pulse is in the middle of two closed curves, which is not shown.) The Gray dotted line is the fitted curve $y=2x+b$ to get the noise susceptibility. The nested graph in (c) is the infidelity of mixed noise RCP with a wider noise amplitude range. (d) The fitted noise susceptibility varies with the width of the noise spectrum.}
  \label{Results_X_03}
\end{figure}

A single qubit subject to longitudinal noise is described by Hamiltonian $H(t) = \tfrac{\delta(t)}{2}\,\hat{Z} \;+\; \tfrac{\Omega(t)}{2}\,\hat{X}$, where $\delta(t)$ denotes noise and $\Omega(t)$ is the control field. Pauli operator $\hat{Z}$ captures the coupling operator for dephasing noise. To implement a $X_\pi$ gate, usually controls on $\hat{X}$ are needed, so the noise coupling strength $c_z$ remains constant and can be included in $\delta(t)=c_z \beta(t)$. The effective noise Hamiltonian reads
\begin{equation}
    \widetilde{H}_n(t) =\tfrac{\delta(t)}{2}\,\bigl[\hat{Z}\cos\phi(t)\;+\;\hat{Y}\,\sin\phi(t)\bigr],
\end{equation}
where $\phi(t)=\int_0^t \Omega(t')\,dt'$.

To quantify the susceptibility of noise across different frequency components, we employ the filter function~\cite{supp}:
\begin{equation}
F(\omega,T) \;=\; 
\Bigl|\int_0^T \sin\phi(t)\,e^{-i\omega t}\,dt\Bigr|^2 \;+\;
\Bigl|\int_0^T \cos\phi(t)\,e^{-i\omega t}\,dt\Bigr|^2.
\end{equation}
Since its analytic form is not strictly necessary for pulse optimization, one can numerically shape $\Omega(t)$ to minimize $F(\omega,T)$ in preselected frequency bands, effectively realizing a multi-band-stop filter for noise. We target two noise bands: low frequencies $(0, \omega_0)$ and high frequencies $(2.5\omega_0, 3.5\omega_0)$, where $\omega_0$ is the Fourier fundamental frequency. The optimized robust control pulse (RCP) achieves $>10\times$ suppression in these bands compared to standard $\pi$ and $9\pi$-cosine pulses [Fig.~\ref{Results_X_03}(b)].

Fig.\ref{Results_X_03} shows the results of numerical verification of $X_\pi$ gates under time-dependent colored noise, implemented using four different waveforms for the control pulses. They are a cos $\pi$-pulse, a cos $9\pi$-pulse, and two robust control pulses (RCP) for static noise ($\text{RCP}_s-\pi$) and multicolored noise ($\text{RCP}_c-9\pi$). As seen in Fig.~\ref{Results_X_03}(b), their filter functions display distinct frequency-domain suppression profiles (shaded regions) alongside the composite noise power spectral density, illustrating the advantage of a band-selective design. At small $A$, infidelity saturates at the noise-free limit ($\sim10^{-5}$), dominated by numerical precision. For moderate $A$, the RCP’s minimized $b_2$ ensures a  slightly shift of infidelity from quadratic scaling, while traditional pulses follow $\propto A^2$. At large $A$, quartic terms ($b_4A^4$) dominate due to nonlinear noise coupling, causing deviation from the $A^2$ fit. The RCP maintains a $10\times$ infidelity reduction over cosine pulses across all regimes, validating its robustness, as corroborated by the noise susceptibility results depicted in Fig.~\ref{Results_X_03}(d).

\paragraph{Robust Two-Qubit Gate.-}We study a robust CZ gate for inductively coupled fluxonium qubits~\cite{ma2024native}, which are sensitive to low-frequency flux noise when detuned from their sweet spots. This task has been identified as a significant challenge~\cite{hai2022universal,ma2024native}. By pulsing the magnetic flux $\Phi(t)$, one obtains an effective dispersive Hamiltonian inducing the CZ phase~\cite{supp}:
\begin{equation}
H_0\,\approx\,\tfrac12\,\Omega(\Phi)\,\hat{Z}_A+\tfrac12\,\eta\,\Omega(\Phi)\,\hat{Z}_B+J_{zz}^{\parallel}(\Phi)\,\hat{Z}_A\hat{Z}_B,
\end{equation}
where $\Omega(\Phi)=\sqrt{(2I_p^j\Phi)^2+\Delta_j^2}-\Delta_j$ denotes the qubit frequency shift, $I_p^j,\Delta_j$ are fluxonium parameters, and $\eta=\Delta_B/\Delta_A$, see Fig.\ref{Results_CZ}(a). $I^j_p$ are stable system parameters in each fluxonium and represent the persistent superconducting loop current depending on their Josephson tunneling energies $\Delta_j$. The coefficient $\eta=\frac{\Delta_B}{\Delta_A}$ is a constant. Since $\Phi(t)$ also enters the noise Hamiltonian $\tilde{H}_n=\beta(t)[\tfrac12\,c_{z_A}(t)\,\hat{Z}_A+\tfrac12\,c_{z_B}(t)\,\hat{Z}_B+c_{zz}(t)\,\hat{Z}_A\hat{Z}_B]$, filter functions are employed to minimize the impact of low-frequency flux noise, the main source of decoherence. For simplicity, we assume these two qubits couple to the same noise environment, $\beta_j(t)=\beta(t)$. Here the terms $c_{z_j}(t)=4\Delta_jI_j^2\Phi(t)/\sqrt{\Delta_j^2+4I_j^2\Phi^2(t)}$ and $c_{zz}(t)=4J\Delta_j^2I_j^2\Phi(t)/(\Delta_j^2+4I_j^2\Phi^2(t))^2$ control the coupling strength to dephasing noise and controlled-phase noise, respectively, as shown in Fig.\ref{Results_CZ}(a) with a rescaling. Numerically, we design the $20$ns smooth robust control pulse (RCP) $\Phi^{\text{RCP}}(t)$ for the CZ gate, Fig.\ref{Results_CZ}(b), with the band-stop filter for the low-frequency noise. Our simulations compare RCP with square, sinusoidal, and net-zero pulses~\cite{negirneac2021high}, revealing that RCP yields two orders of magnitude lower gate infidelity across various noise amplitudes and bandwidths, see Fig.\ref{Results_CZ}(d). The comparison of their filter functions $F(\omega,T)=\sum_{i}\left|\int_0^Tc_{i}(t)e^{-i\omega t}dt\right|^2$ is shown in  Fig.\ref{Results_CZ}(c), $i=z_A,z_B,zz$, which shows at least two orders of suppression at the low-frequency noise region (gray) using RCP. Hence, RCP's infidelity is improved by more than two orders of magnitude compared to trivial pulses, as shown in Fig.\ref{Results_CZ}(d). Fig.\ref{Results_CZ}(e) further shows that RCP has better noise susceptibility as the low-frequency noise's spectral bandwidth increases. Notably, this noise susceptibility doesn't depend on a specific gate, but only reflects the robustness of the control pulse. The parameters  $\Delta_{A(B)}/2\pi=1(1.3)\mathrm{GHz}$, $I_p^{A(B)}=10\Delta_{A(B)}/\Phi_0$ are used in the numerical study to fit the experimental data. 

\begin{figure}[t]
  \centering
  \includegraphics[width=1\linewidth]{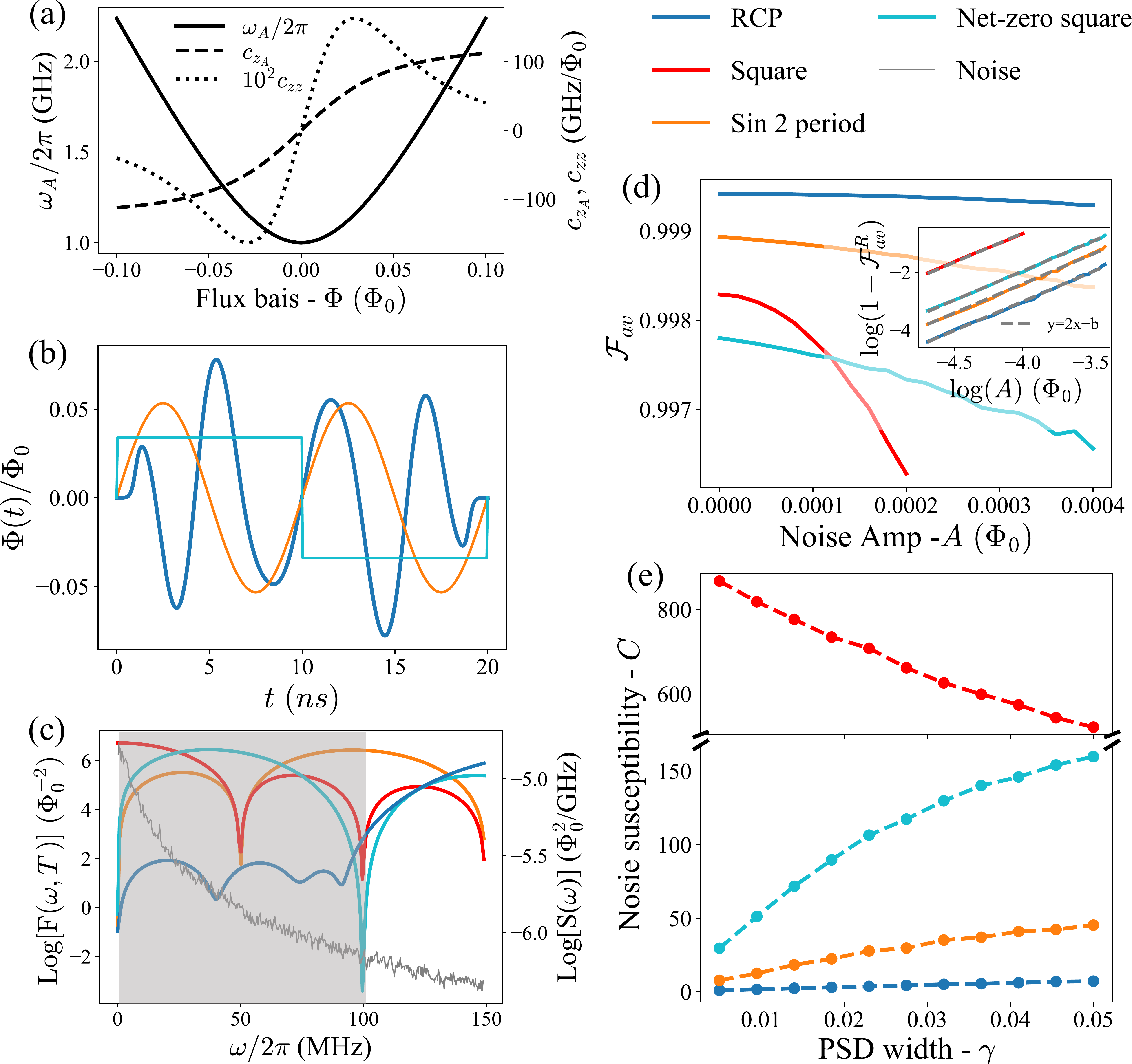}
  \caption{Comparison of CZ gates in the presence of low-frequency noise. (a) Dependence of Qubit~A's frequency and its noise-coupling strengths $c_{z_A}, c_{zz}$ on the applied magnetic flux $\Phi(t)$. (b) Waveforms of the RCP (blue solid) and peak amplitudes (dashed lines) for sine/cosine-modulated pulses. (c) Filter functions of different pulse shapes with shaded regions denoting targeted suppression bands. (d) Average CZ-gate fidelity versus amplitude of the time-dependent noise for the RCP, standard square pulse, a net-zero pulse with square function and sine function. The nested log-scale inset shows the gate infidelity, with a gray dotted line fitting $y=2x + b$ to quantify noise susceptibility. (e) Noise susceptibility $C$ as a function of the PSD bandwidth $\gamma$. Values are normalized to the RCP’s minimum susceptibility. Data points (dots) are plotted by fitting the infidelities from the inset in (d).}
  \label{Results_CZ}
\end{figure}

\paragraph{Enhanced Sensing of AC signals.-}
We demonstrate an extension of our filtering framework to AC signal sensing using flux qubits. The system Hamiltonian is $H(t)= \frac{\omega(\Phi(t))}{2}\hat{Z}$, where total flux $\Phi(t)$ includes bias $\Phi_b(t)$, noise $\beta(t)$, and signal $\Phi_s(t)=B\cos(\omega_s t+\alpha)$. For $\Phi_b(t)\gg\Phi_s(t),\beta(t)$, fluctuations can be approximated as $\delta\omega_n(t) = c_{z}(t)\beta(t)$ and $\delta\omega_s(t) =B c_z(t)\cos(\omega_s t+\alpha)$, where $c_z(t)=\frac{4\Delta I_p^2 \Phi(t)}{\sqrt{\Delta^2+4I_p^2\Phi^2(t)}}$. Starting from the initial state $\lvert + \rangle = \frac{1}{\sqrt{2}}(\lvert 0 \rangle + \lvert 1 \rangle)$, the measurement probabilities of $\lvert \pm \rangle$ under multicolored Gaussian noise are 
\begin{equation}
\langle P_{\pm}(B)\rangle = \frac{1}{2}(1\mp e^{-\chi(t)/2}\cos(B\Gamma(t))),
\end{equation}
where $\Gamma(t)=\int_0^t c_z(v)\cos(\omega_s v+\alpha)dv$ and $\chi(t)=\int_{-\infty}^{\infty}S(\omega)F(\omega,t)d\omega$ with filter function $F(\omega,t)=|c_z(\omega,t)|^2$. The signal amplitude $B$ can be determined by measuring the Ramsey oscillation frequency. For periodic control with period $T$ satisfying $\omega_sT=2k\pi$, this frequency equals $\bar{\Gamma}B$ where $\bar{\Gamma}=\Gamma(T)/T$. The corresponding Fisher information takes the form:
\begin{equation}
\mathcal{I}(B)=\sum_n \frac{1}{1 + \frac{1/e^{-2\chi(nT)} - 1}{\sin^2(B\bar{\Gamma}nT)}}(\bar{\Gamma}nT)^2
\end{equation}
The equation elucidates our optimization strategy: to augment the sensitivity to unknown field strength $B$, it is imperative to maximize $\bar{\Gamma}$ (thereby enhancing signal sensitivity) while concurrently minimizing $\chi(t)$ (thereby attenuating noise effects). As demonstrated in \cite{supp}, the objective now becomes to maximize $\int_{\sum_i\mathcal{B}_i}F(\omega,T)\,d\omega$ within signal bands $\sum_i\mathcal{B}_i$ and minimize $\int_{\sum_j\mathcal{A}_j}F(\omega,T)\,d\omega$ within noise bands $\sum_i\mathcal{A}_i$. This approach directly translates into the filter function design principles expounded upon earlier. 

\begin{figure}[t]
  \centering
  \includegraphics[width=1\linewidth]{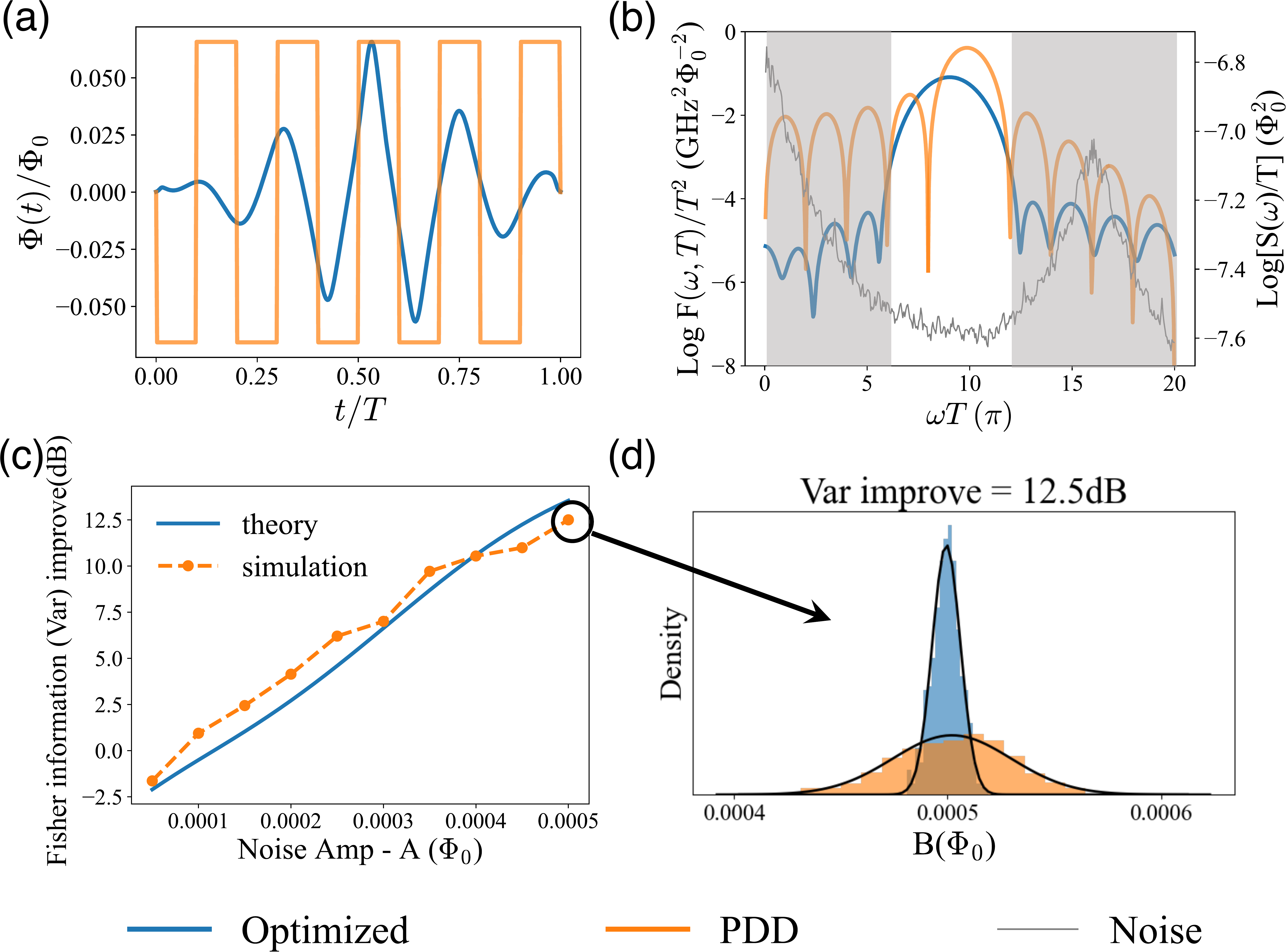}
  \caption{Results of the optimized sensing. (a) Flux bias $\Phi(t)$ for the optimized control and PDD. (b) The corresponding filter functions in a logarithmic scale, highlighting noise-suppression (shaded) and signal-amplification (unshaded) regions.  (c) the improvement in the variance of the simulation results of the optimized waveform versus PDD as noise amplitude $A$ increases, as well as the improvement in the theoretical fisher information. (d) The Gaussian distribution of the measured $B$ values under each control scheme.}
  \label{Results_sensing}
\end{figure}

To demonstrate the efficacy of our approach, we designed an optimized flux-control waveform targeting a signal band of $(\omega_{inf},\omega_{uv-})$ while suppressing low-frequency noise in the range of $(0,\omega_{inf})$ and high-frequency noise in the range of $(\omega_{uv-},\omega_{uv+})$. Here, the control pulse period  $T$ is tuned to fit the signal band $T=\frac{2\pi k}{\omega_{uv-}-\omega_{inf}}$, where $k$ determines the frequency resolution of detection. In our numerical example, we choose $\{\omega_{inf}, \omega_{uv-}, \omega_{uv-} \}=\{6, 12, 20\}\pi/T$ with the detection resolution $1/T=10^7$ Hz. The Lorentzian noise peaks at $0$ and $16\pi/T$ (see the gray curve in Fig.\ref{Results_sensing}(b)). Figure~\ref{Results_sensing}(a) shows one period of  the optimized waveform compared to a modified periodic dynamical decoupling (PDD) sequence. The corresponding filter functions in Fig.~\ref{Results_sensing}(b) clearly demonstrate our waveform's enhanced spectral selectivity. For each estimate, we aggregated 300 single-shot measurements and repeated this process 600 times to establish statistical significance. Figure~\ref{Results_sensing}(c) quantitatively illustrates the enhancement in performance correlated with increasing noise amplitude, demonstrating that our method achieves an improvement of $4$ to $10$ dB in estimation variance relative to PDD. The distribution of estimated $B$ values in Fig.~\ref{Results_sensing}(d) for noise amplitude $A=0.0005\Phi_0$ further illustrates this improvement. These results demonstrate that carefully engineered flux-control waveforms can significantly enhance quantum sensing precision in noisy environments, providing a practical advantage over conventional techniques. Our approach is particularly valuable for applications requiring frequency-selective measurements in the presence of structured noise backgrounds. For example, this enhanced control could be the first stage of signal amplification in QND of quantum states before sending to signal amplifiers such as flux-driven Josephson parametric amplifier~\cite{yamamoto2008flux}. The full derivation and analysis of our sensing protocol are provided in the Supplemental Material\cite{supp}.

\paragraph{Conclusion.-}This work presents a generalized filter-function framework that concurrently suppresses noise and amplifies signals across programmable frequency bands, offering a universal tool to tackle spectrally complex noise while enhancing precision in quantum control and sensing. Integrating the framework with the COCOA algorithm leads to an efficient way to engineer smooth, hardware-compatible waveforms that ensure broad applicability across physical platforms. By unifying noise suppression and signal enhancement, this framework bridges quantum control and sensing, transforming spectral vulnerabilities into programmable assets. Our findings contribute to robust quantum operations and highlight the filter function formalism as a powerful tool for processing correlated and multi-colored noise in quantum devices.

\paragraph{Acknowledgement.} We thank fruitful discussions with Xiaoting Wang, Haidong Yuan and Yuanlong Wang. This work is supported by the Key-Area Research and Development Program of Guang-Dong Province (Grant No. 2018B030326001), the Science, Technology and Innovation Commission of Shenzhen Municipality (JCYJ20170412152620376, KYTDPT20181011104202253), and the Shenzhen Science and Technology Program (KQTD20200820113010023). 

\bibliography{refs}
\bibliographystyle{apsrev4-2}

\newpage

   \setcounter{figure}{0}
    \setcounter{table}{0}
    \setcounter{equation}{0}
\renewcommand{\theequation}{S\arabic{equation}}
    \renewcommand{\thefigure}{S\arabic{figure}}
    \renewcommand{\thetable}{S\arabic{table}}
\graphicspath{{figures_supple/}}

\begin{titlepage}
\begin{center}
\large\textbf{Supplementary for "Enhanced Quantum Signal Control and Sensing Under Multicolored Noise via Generalized Filter Function Framework"}

\vspace{1cm}
\normalsize
Zhi-Da Zhang$^{1,2}$, Yao Song$^{1,2}$, Wen-Zheng Dong$^{1}$, Xiu-Hao Deng$^{1,2,*}$

\vspace{0.5cm}
\small
$^{1}$Shenzhen International Quantum Academy (SIQA), Futian District, Shenzhen, P. R. China\\
$^{2}$Shenzhen Institute for Quantum Science and Engineering (SIQSE), Southern University of Science and Technology, Shenzhen, P. R. China\\
$^{*}$Email: dengxh@sustech.edu.cn
\end{center}
\end{titlepage}

\appendix
\onecolumngrid

\section{extended filter function framework} \label{sec: ff method}

First, we introduce the general extend filter function framework for semiclassical noise sources, which can be extended to quantum noise\cite{paz2014general}.  Consider an open multi-qubit system with the Hamiltonian: $H=H_s+H_{ctrl}+H_n \equiv H_0+H_n$,  where $H_s,H_{ctrl},H_n$ represent the intrinsic system Hamiltonian, the control Hamiltonian, and the noise Hamiltonian, respectively. In the operator basis $ \{\Lambda_u\} $ typically chosen to be tensor products of Pauli operators, $H$ takes the form:
\begin{equation}
H(t) = \sum_u h_u(t)\Lambda_u+\sum_u\delta_u(t)\Lambda_u
\end{equation}
The first term is the noiseless part $H_0$, and the second term is the noise part $H_n$.  Taking the evolution  of $H_0$ as a reference, the total evolution takes the form:  
\begin{equation}
U(t)=U_0(t) U_e(t)
\end{equation}
where $U_0$ represents the noiseless evolution determined by $H_0$, and $U_e$ describes the \textit{error evolution operator}~\cite{hai2022universal} generated by $H_n$. Generally,
\begin{equation}
U_{0}(t)=\mathcal{T}_{+}e^{-i\int_{0}^{t}dsH_{0}(s)},\quad U_e(t)=\mathcal{T}_{+}e^{-i\int_{0}^{t}ds\widetilde{H}_n(s)},
\end{equation}
with $ \widetilde{H}_n = U_0^\dagger H_n U_0 $ is the noise interaction Hamiltonian, which can be generalized as:
\begin{equation}
\widetilde{H}_n(t) \equiv \sum_{u, v} y_{u, v}(t) \, \delta_u(t) \, \Lambda_{v}
\end{equation}
where $ y_{u, v}(t) = \mathrm{Tr}_{S}[U_{0}^{\dagger}(t) \Lambda_{u} U_{0}(t) \Lambda_{v}] / d $, with $ d = 2^n $ as the system dimension.

The effective noise strength $\delta_u(t)$ encountered by the quantum system is associated with its physical origin $ \beta_u(t) $ through the coupling strength $ c_u(t) $, as expressed by $\delta_u(t) = c_u(t) \beta_u(t)$.  For instance, $ \beta(t) $ could represent charge or flux noise from the environment, while $ \delta(t) $ denotes the noise amplitude affecting the system. 
The \textit{error evolution operator}  $U_e$ can be expressed using the Magnus expansion as $U_e = e^{\sum \Omega_\alpha}$ , where the $\alpha$th order Magnus term $\Omega_\alpha$ is a time-ordered integral of $\widetilde{H}_n(t)$, over the volume $V_{\alpha}\equiv\{0\leq t_{\alpha}\leq t_{\alpha-1}\leq\ldots\leq t_{2}\leq t_{1}\leq T\}$
\begin{equation}
\Omega_{\alpha}=\sum_{\vec{u},\vec{v}}\int_{V_{\alpha}}d\vec{t} ^{\alpha}\sum_{p\epsilon\Pi[\{t_{j}\}]}f^\alpha(p)\  y^{[\alpha]}_{\vec{u},\vec{v}}(p) \beta^{[\alpha]}_{\vec{u}}(p) \Lambda^{[\alpha]}_{\vec{v}}
\end{equation}
with
\begin{equation}
\begin{aligned}
    &d\vec{t} ^{\alpha}= dt_1dt_2...dt_\alpha,\\
    &y^{[\alpha]}_{\vec{u},\vec{v}} (p)\equiv y_{u_1v_1}(p(t_1))\cdotp\cdotp\cdotp 
    y_{u_\alpha v_\alpha}(p(t_\alpha)), \\
    &\beta^{[\alpha]}_{\vec{u}}(p) \equiv \beta_{u_1}(p(t_1))\cdotp\cdotp\cdotp \beta_{u_\alpha }(p(t_\alpha)),\\
    &\Lambda^{[\alpha]}_{\vec{v}}=\Lambda_{v_1}\cdotp\cdotp\cdotp \Lambda_{v_\alpha},
\end{aligned}
\end{equation}
and $f^\alpha(p)$ are functions representing the coefficients, which contain the imaginary number $i$, of the $\alpha$th order Magnus expansion and signs for different permutations.

To quantify the impact of noise, we calculate the average gate fidelity~\cite{nielsen2002simple}, given by
\begin{equation}\label{fid_av}
\mathcal{F}_{av}=\langle|\frac1d\ \mathrm{Tr}[U_G^\dagger U(\tau)]|^2\rangle=\frac1{d^2}\langle|\mathrm{Tr}[U_e(\tau)]|^2\rangle 
\end{equation}
Here, $U_G$ is the target gate, and we assume that $U_0(\tau) = U_G$, and $\langle\cdot\rangle$ represents the average over semi-classical noise. Taking the Taylor expansion to $U_e$ and using the relationship $|\mathrm{Tr}[U_e(\tau)]|^2=\mathrm{Tr}[U_e(\tau)]\cdot\mathrm{Tr}[U_e(\tau)^\dagger]$, we obtain: 
\begin{equation}\label{whole_fid_S}
\begin{aligned}
\mathcal{F}_{avg}&= \sum_{r,r'}\frac{1}{r!r'!} \cdot \frac1{d^2} \mathrm{Tr}[(\sum \Omega_\alpha)^r] \mathrm{Tr}[(\sum \Omega^\dagger_\alpha)^{r'}]\\
&= \sum_{}\frac{1}{r!r'!(2\pi)^{||\vec{\alpha}_r||_1+||\vec{\alpha}'_{r'}||_1}} \mathcal{O}^{[r,r']}.
\end{aligned}
\end{equation}
The final expression involves multiple summations on $r,r';\ \vec{\alpha}_r,\vec{\alpha}_r';\ \vec{u}^{\alpha_1}\cdotp\cdotp\cdotp\vec{u}^{\alpha_r},\vec{v} ^{\alpha_1}\cdotp\cdotp\cdotp\vec{v}^{\alpha_r},\vec{u}^{\alpha'_1}\cdotp\cdotp\cdotp\vec{u}^{\alpha'_{r'}},\vec{v} ^{\alpha'_1}\cdotp\cdotp\cdotp\vec{v}^{\alpha'_{r'}}$. The dimension of $\vec{\alpha}_r$ is decided by $r$, and the dimension of $\vec{u}^{\alpha_j},\vec{v}^{\alpha_j}$ is decided by the corresponding $\alpha_j$. The $2\pi$ term comes from Fourier transformation. Here, $\mathcal{O}^{[r,r']}$ takes the form:

\begin{equation}
\begin{aligned} 
\mathcal{O}^{[r,r']}=&\int D\vec{\omega}\  G^{[\alpha_1]}\cdotp\cdotp\cdotp G^{[\alpha_r]}G^{[\alpha'_1]}\cdotp\cdotp\cdotp G^{[\alpha'_{r'}]}S(\vec{\omega}^{\alpha_1},\cdotp\cdotp\cdotp ,\vec{\omega}^{\alpha_r},\vec{\omega}^{\alpha'_1},\cdotp\cdotp\cdotp ,\vec{\omega}^{\alpha'_{r'}})\frac{1}{d^2}\mathrm{Tr}(\Lambda^{[\alpha_1]}\cdotp\cdotp\cdotp \Lambda^{[\alpha_r]})\mathrm{Tr}(\Lambda^{[\alpha'_1]}\cdotp\cdotp\cdotp \Lambda^{[\alpha'_{r'}]})\\
=&\int D\vec{\omega}\ G(\vec{\omega}^{\alpha_1}\cdotp\cdotp\cdotp \vec{\omega}^{\alpha_r}\vec{\omega}^{\alpha'_1}\cdotp\cdotp\cdotp \vec{\omega}^{\alpha'_{r'}},T) S(\vec{\omega}^{\alpha_1},\cdotp\cdotp\cdotp ,\vec{\omega}^{\alpha_r},\vec{\omega}^{\alpha'_1},\cdotp\cdotp\cdotp ,\vec{\omega}^{\alpha'_{r'}})\frac{1}{d^2}\mathrm{Tr}(\Lambda^{[\alpha_1]}\cdotp\cdotp\cdotp \Lambda^{[\alpha_r]})\mathrm{Tr}(\Lambda^{[\alpha'_1]}\cdotp\cdotp\cdotp \Lambda^{[\alpha'_{r'}]})
\end{aligned}
\end{equation}

where $G^{[\alpha]}= G^{[\alpha]}_{\vec{u},\vec{v}}(\vec{\omega}^\alpha, T)$  is the $\alpha$-th order subfilter function:
\begin{equation}\label{sub_filter}
G^{[\alpha]}_{\vec{u},\vec{v}}(\vec{\omega}^\alpha, T) = \int_{V_\alpha} d\vec{t}^\alpha \sum_{p \in \Pi[\{t_j\}]} f^\alpha(p) \, y^{[\alpha]}_{\vec{u},\vec{v}}(p)e^{i\vec{\omega}^\alpha \cdot p(\vec{t}^\alpha)}.
\end{equation}
Their product \( G(\vec{\omega}^{\alpha_1} \cdots \vec{\omega}^{\alpha_r} \vec{\omega}^{\alpha'_1} \cdots \vec{\omega}^{\alpha'_{r'}}, T) \) represents the general filter function for the noise spectral density \( S \), which is defined as:
\begin{equation}
    S(\vec{\omega}^{\alpha_1}, \vec{\omega}^{\alpha_2}, \dots) = \int D\vec{t} \langle \beta^{[\alpha_1]}_{\vec{j}^{\alpha_1}}(\vec{t}^{\alpha_1}) \beta^{[\alpha_2]}_{\vec{j}^{\alpha_2}}(\vec{t}^{\alpha_2}) \dots \rangle e^{-i(\vec{\omega}^{\alpha_1} \cdot \vec{t}^{\alpha_1} + \vec{\omega}^{\alpha_2} \cdot \vec{t}^{\alpha_2} + \dots)}.
\end{equation}

The term \( \frac{1}{d^2} \mathrm{Tr}(\cdots) \) acts as a weight factor, which can be 0 or 1.

The symbols $\int D\vec{t},\int D\vec{\omega}$ represent the multidimensional  integrals over all the relevant $\vec{t}^\alpha,\vec{\omega}^\alpha$.

Instead of the Taylor expansion order $r,r'$, we can also sort $\mathcal{F}_{av}$ in terms of noise order, which has further complication. But for small noise amplitudes, we can truncate Eq.~\eqref{whole_fid_S} to the second noise order.
\begin{equation}
\begin{aligned}
\mathcal{F}_{av}\approx&\frac1{d^2}[\mathrm{Tr(I)Tr(I)}+\mathrm{Tr(\Omega_1)Tr(I)}+\mathrm{Tr(I)Tr(\Omega_1^\dagger)}+ \mathrm{Tr(\Omega_2)Tr(I)}+ \mathrm{Tr(I)Tr(\Omega_2^\dagger)}\\
&+\mathrm{Tr(\Omega_1)Tr(\Omega_1^\dagger)}+\frac12\mathrm{Tr(I)Tr(\Omega_1\Omega_1)} +\frac12\mathrm{Tr(\Omega_1^\dagger\Omega_1^\dagger)Tr(I)}]\\
=&1+\frac1d\mathrm{Tr(\Omega_1\Omega_1)}\\
=&1+\sum_{u_1,u_2,v_1=v_2} \int D\vec{\omega}\ \frac{1}{(2\pi)^2}G(\omega_1,\omega_2,T) S(\omega_1,\omega_2)
\end{aligned}
\end{equation}
Simplification uses the relation $\mathrm{Tr}[\Omega_1]=0$, $ \Omega_\alpha^\dagger = -\Omega_\alpha $ and rewrites the symbol $(\vec{\omega}^{\alpha_1},\vec{\omega}^{\alpha_2})$ as $(\omega_1,\omega_2)$, since $\alpha_1=\alpha_2=1$. Similar notation changes apply to $\vec{u},\vec{v}$.

Assuming noise in different directions $u$ is uncorrelated, such that $S_{u_1,u_2}(\omega_1,\omega_2)=0,u_1\neq u_2$, and treating the noise source as a Gaussian stationary process, with $S(\omega_1,\omega_2)=2\pi S_u(\omega_1)\delta(\omega_1+\omega_2)$, we further simplify the result as:
\begin{equation}\label{approx_fid_S}
\begin{aligned}
\mathcal{F}_{av}&\approx1+\sum_{u,v} \int \frac{d\omega }{2\pi}G_{u,v}(\omega,-\omega,T) S_u(\omega)\\
&=1-\sum_{u,v} \int \frac{d\omega }{2\pi}|G^{[1]}_{u,v}(\omega,T)|^2 S_u(\omega)\\
&=1-\sum_u \int \frac{d\omega }{2\pi} F_u(\omega,T)S_u(\omega)
\end{aligned}
\end{equation}
Where $F_u(\omega,T)=\sum_{v} |G^{[1]}_{u,v}(\omega,T)|^2$ is the second order filter function for noise $\beta_u$($\delta_u$). Suppressing \( F_u(\omega, T) \) leads to robust quantum gates, while amplifying specific frequency ranges enables high signal-to-noise ratio sensing.

\section{Optimization Method}\label{app_opt method}
We decompose the robust waveform optimization problem into two primary objectives. The first objective is to maximize the ideal gate fidelity by ensuring that $ U_0(T) = U_G $, while the second objective is to minimize the effects of noise by driving $ U_e(T) \to \mathrm{I} $. The combined optimization cost function can thus be formulated as:
\begin{equation}
F_{cost}(T) =  w_1\mathcal{F}_{G}(T)+w_2L(T)
\end{equation}
where:
\begin{equation}
\begin{aligned}
&\mathcal{F}_G(T)=\frac{1}{d}|\mathrm{Tr}[U_G^{\dagger}U_0(T)]|^2\\
&\Gamma(T)=\int_{\sum_{\mathcal{A}_i}}F(\omega)d\omega
\end{aligned}
\end{equation}
Here, $ \mathcal{A}_i $ represents a specific frequency band selected based on the noise spectrum, and the parameters $ w_1 $ and $ w_2 $ are weight factors that balance the trade-off between these two competing objectives.

The cost function of enhanced sensing also has two primary objectives. One is to maximize the filter function in the signal frequency ranges to improve the signal susceptibility. The other is to minimize the filter function in noise frequency ranges to reduce the impact of decoherence. The combined optimization cost function takes the form:
\begin{equation}
F_{cost}(T) =  w_1L_1(T)+w_2L_2(T)
\end{equation}
where:
\begin{equation}
\begin{aligned}
&L_1(T)=\int_{\sum_{\mathcal{A}_i}}F(\omega)d\omega\\
&L_2(T)=-\int_{\sum_{\mathcal{B}_i}}F(\omega)d\omega
\end{aligned}
\end{equation}
The definition of $\mathcal{A}_i$ is the same as above. $\mathcal{B}_i$ represents a specific frequency band covering the signal frequency distribution. There are still trade-off weight factors between these two competing objectives.

We employ both cost functions in the COCOA (Control Optimization with Constrained Auto-differentiation) algorithm~\cite{song2022optimizing} to optimize the control pulses. COCOA is conceptually similar to the GRAPE (Gradient Ascent Pulse Engineering) algorithm in that it uses a piecewise constant parametrization of the control waveform. However, COCOA reshapes each resulting piecewise waveform through a post-processing step. The reshaped waveform takes the form:
\begin{equation}
\Omega_0(a_j,\phi_j;t)=\sin\biggl(\frac{\pi t}T\biggr)\biggl(a_0+\sum_{j=1}^na_j\cos\biggl(\frac{2\pi j}Tt+\phi_j\biggr)\biggr)
\end{equation}
Where $a_j,\phi_j$ are the discrete Fourier transform coefficients of the input PWC pulse.  The shaping process is as follows. First, we perform a low-frequency filter to the input waveform, retaining only the initial pulse's $n$ lowest-order Fourier basis terms. This step ensures the simplicity and smoothness of the waveform. Next, we multiply the entire waveform by a half-cycle sine function, ensuring that the waveform starts and ends at zero. While the sine function is commonly used, other functions, such as the sigmoid function, can also be applied to impose constraints on the starting and ending points of the waveform. The initial parameters of the PWC ansatz are picked from Cos or Square pulse. The optimized pulses retain up to seven Fourier components.

\begin{table}[tb]
\centering
\renewcommand{\arraystretch}{1.5} 
\begin{tabular}{@{}ll@{}}

\hline\hline
\textbf{Symbol} & \textbf{Meaning} \\ 
\hline
$k$ & Time slice index ($1, \ldots, N$) \\ 
$\Omega[k]$ & Control amplitude in time slice $k$ \\ 
$F_{cost}$& Cost function\\ 
$\Omega^r[k]$& Control amplitude at iteration $r$\\ 
$N_c$ & Number of Fourier components kept in optimized pulse \\ 
$\alpha$ & Learning rate \\ 
$\epsilon_0$ & Tolerance of cost function \\ 
$U_{G}$& Target operator \\ 
$\Omega^*[k]$& Optimized control amplitude in time slice $k$\\ 
\hline\hline
\end{tabular}
\caption{List of symbols used in the optimization algorithm.}
\label{Tab1}
\end{table}

\begin{algorithm}[tb]
\label{algorithm1}
\caption{Optimization Algorithm based on COCOA}
\KwIn {Cost function: $F_{cost}$; piecewise constant waveform: $\{\Omega^0[k]\}$; iteration parameters: $N_{c}$, $\alpha$, $\epsilon_{0}$.}
\KwOut{Optimized piecewise constant waveform: $\{\bar{\Omega}^*[k]\}$.} 
\BlankLine
\Repeat{$F_{cost} < \epsilon_0$}{
    apply pulse filter: $\{\Omega[k]\} \rightarrow \{\tilde{\Omega}[k]\}$\;
    
    apply pulse constrains: $\{\tilde{\Omega}[k]\} \rightarrow \{\bar{\Omega}[k]\}$\;
    
    solve $i\hbar\frac{\partial U(t)}{\partial t} = H(\Omega(t)) U(t)$ by Runge–Kutta methods (option)\; 

    calculate gate fidelity of $U_G$ (option) \;
    
    calculate the filter function getting $F_{cost}$\;
    
    compute gradient $\{\frac{\partial F_{cost} }{ \partial \Omega[k]}\}$ using auto-differentiation \;

    update $\{\Omega[k]\}$ \;
    
}
reshape optimized result $\{\Omega^*[k]\} \rightarrow \{\tilde{\Omega}^*[k]\} \rightarrow \{\bar{\Omega}^*[k]\}$

\Return{$\{\tilde{\Omega}^*[k]\}$}\;
\end{algorithm}

We use the Adam algorithm for optimization. This is because Adam computes the gradient through automatic differentiation, which eliminates the need to derive the analytical expression for the gradient of the cost function with respect to the waveform parameters. The pseudocode of this algorithm is shown in Algorithm \ref{algorithm1} with the symbol meaning shown in TABLE \ref{Tab1}.

\section{Noise Models}
\label{App_NoiseModel}
In this article, two different methods are used to generate time-domain noise signals. One is to simulate the random telegraph noise (RTN) process. This method well explains the noise characteristics observed in experiments. A single RTN process exhibits a Lorentzian power spectrum. By combining multiple RTN processes, we can construct a $1/f$ noise spectrum. The noise in the time domain can be written as:
\begin{equation}
\delta\beta(t)=\sum_{i=0}^Nw_i\eta_i(t),
\end{equation}
where $\eta_i(t)$ is a single RTN process with characteristic time $\tau_i$, and $w_i$ is a weight factor. The noise power spectrum for $\delta\beta(t)$ can be written as:
\begin{equation}
S(\omega)=\sum_{i=0}^N\frac{4\tau_i}{4+\omega^2\tau_i^2}w_i^2
\end{equation}
This is a sum of Lorentz-type noise spectra. To construct $1/f$  spectra, the parameters $\tau_i,w_i$  are chosen as follows. Here, $\tau_{max},~N$ should be  as large as possible, while $\tau_{min}$ should be as small as possible.
\begin{equation}
\tau_i=\frac{\mathrm{i}}N(\tau_{\max}-\tau_{\min})+\tau_{\min}, \ w_i^2=\frac{A^2}{N\pi}\frac{\tau_{\max}-\tau_{\min}}{\tau_i}
\end{equation}

The second model we use to generate time-dependent noise is the inverse Fourier transform method, which is a mathematically rigorous tool for simulating time-dependent noise with a specified spectrum. By defining the desired power spectral density in the frequency domain, this approach enables the generation of time-domain signals that exhibit the corresponding noise characteristics. While the method is capable of simulating a broad range of spectra, practical considerations such as sampling rates, numerical accuracy, and physical constraints must be accounted for. To ensure compliance with the Gaussian stationary noise condition, a random uniform distribution is applied to the phase. Consequently, the noise in the time domain can be expressed as
\begin{equation}
\beta(t)=\sum_iF_i\cos(\omega_it+\phi_i)
\end{equation}
where  $F_i^2(\omega_i) = \frac{2}{\pi}\int\limits_{\omega_i-\delta\omega/2}^{\omega_i+\delta\omega/2}S(\omega)d\omega$ , $\delta\omega = \omega_i-\omega_{i-1}$. $\phi_i$ is a uniform distribution on $[0,2\pi]$. 

\section{Robust single-qubit gate}
\label{SingleQBG}
In this section, we illustrate how to realize a noise-resilient single-qubit $X_\pi$ gate under both low- and high-frequency noise. By optimizing a carefully chosen control waveform, we suppress noise over multiple frequency ranges to achieve high gate fidelity.
A single qubit subject to longitudinal noise typically follows the Hamiltonian $H(t) = \tfrac{\delta(t)}{2}\,\hat{Z} \;+\; \tfrac{\Omega(t)}{2}\,\hat{X}$, where $\delta(t)$ denotes noise and $\Omega(t)$ is the control field. To implement a $X_\pi$ gate, usually controls on $\hat{X}$ are needed, so the noise coupling strength $c_z$ remains constant and can be included in $\delta(t)=c_z \beta(t)$. During an $X_\pi$ rotation, the control unitary evolution takes the form:
\begin{equation}
    U_c=e^{-i\int \frac{\Omega(v)}{2}\hat{X} dv}
\end{equation}
The noise Hamiltonian in the interaction picture of $U_c$ transforms as:
\begin{equation}
\widetilde{H}_n(t) \;=\; \tfrac{\delta(t)}{2}\,U_c^\dagger\,\hat{Z}\,U_c 
\;=\; \tfrac{\delta(t)}{2}\,\bigl[\hat{Z}\cos\phi(t)\;+\;\sigma_y\,\sin\phi(t)\bigr],
\end{equation}
where $\phi(t)=\int_0^t \Omega(t')\,dt'$.  The average gate fidelity Eq.\ref{approx_fid_S} in here is:
\begin{equation}\label{approx_fid_S_X}
\mathcal{F}_{avg}\approx1- \int \frac{d\omega }{2\pi} F(\omega,T)S(\omega)
\end{equation}
The filter function is calculated by Eq.\ref{sub_filter} with $y_{zz}(t)=\cos\phi(t),y_{zy}(t)=\sin\phi(t)$
\begin{equation}
\begin{aligned}
F(\omega,T)&=\sum_{v=z,y} |G^{[1]}_{v}(\omega,T)|^2 =|\int\limits_0^T\sin\phi(t)e^{-i\omega t}dt|^2+|\int\limits_0^T\cos\phi(t)e^{-i\omega t}dt|^2
\end{aligned}
\end{equation}
 In practice, calculating the analytical form of the filter function is not necessary for the optimization process. However, it is valuable for verifying numerical results.

By strategically shaping $\Omega(t)$, one can suppress $F(\omega,T)$ in chosen frequency intervals, thereby enhancing gate robustness. We take three different frequency band selection strategies for optimization, targeting low-frequency noise, high-frequency noise, and a scenario where both types coexist.  For the low-frequency and high-frequency noise cases, we optimized a single-frequency band. In the mixed noise case, we optimized two frequency bands simultaneously. 

\begin{figure}
  \centering
  \includegraphics[width=0.9\linewidth]{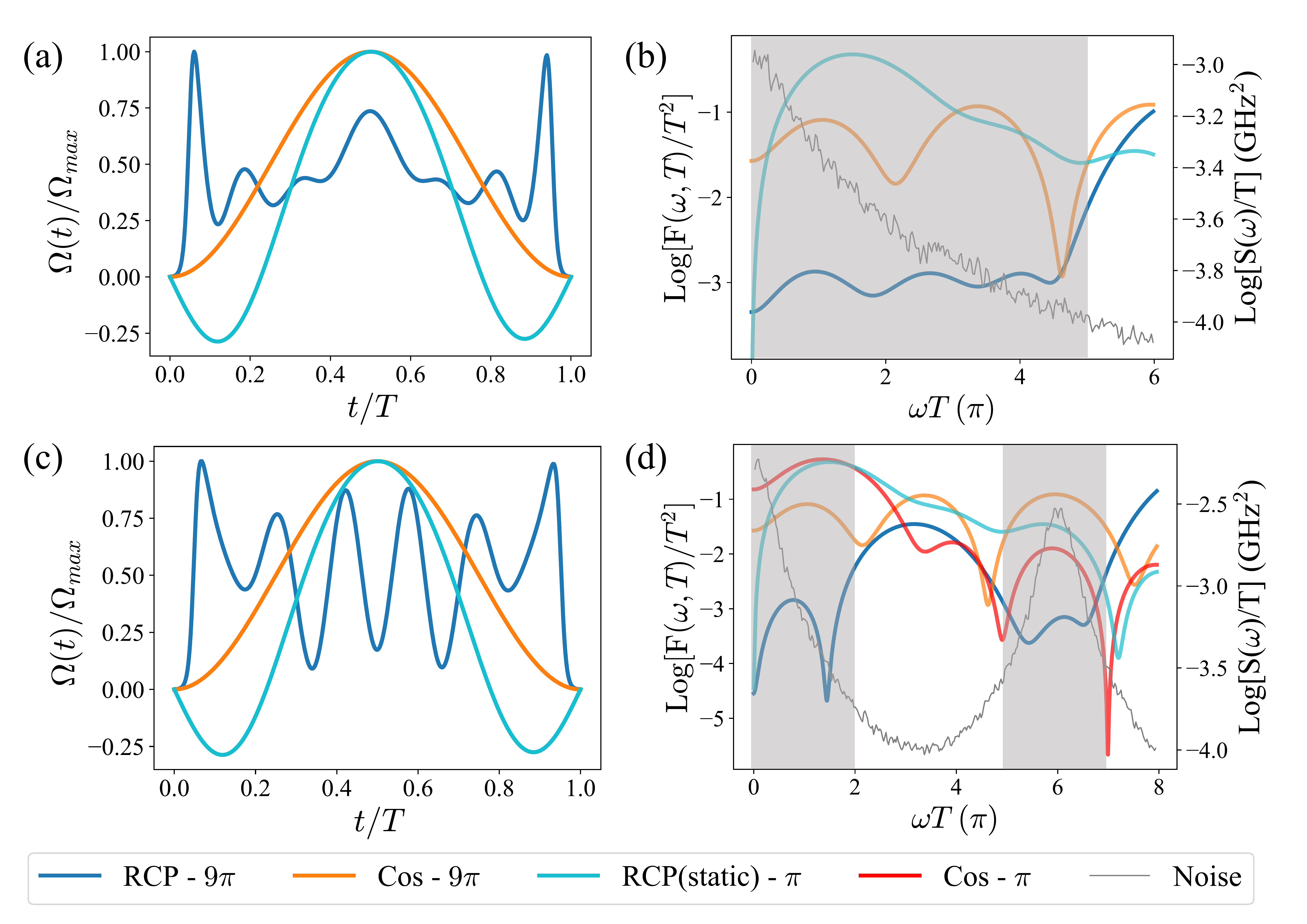}
  \caption{Optimized results of X gates for low-frequency and mixed noise. (a), (c) The pulse waveforms of  Cos -pulse, static RCP, and the RCP for low-frequency(a) and mixed (c) noise. (b), (d) is the corresponding filter function for (a), (c). The shadow areas indicate the target optimization bands and the simulation results of the corresponding noise spectrum are drawn in the filter function diagram. }
  \label{FilterX_0+03}
\end{figure}

First, consider the scenario where the noise power spectrum is concentrated near zero, which is the most common case. This type of noise spectrum is typically of the form $\frac1{f^\alpha}$ or Lorentz-type. The optimal filtering band is $(0,\omega_{1})$, where a larger $\omega_{1}$ results in better filtering. However, in practice, the amplitude and complexity of the waveform limit the maximum value of $ \omega_{1} $. For instance, Fig.\ref{FilterX_0+03}(a) demonstrates the optimized robust control pulse(RCP) with $\omega_{1} = 2.5\omega_0, \omega_0 = 2\pi/T$ is the fundamental Fourier transformation frequency.  We compare it with the geometric gate operation for quasi-static(static RCP) noise\cite{hai2022universal} and the commonly used cosine waveform, specifically the Cos $9\pi$-pulse. The Cos $9\pi$-pulse was chosen because the robust pulse gives a rotation of 9$\pi$, and we need to show the advantages of the optimized pulse at the same control cost. The filter functions of these waveforms are shown in Fig.\ref{FilterX_0+03}(b), we can see the optimized waveform achieves at least two orders of magnitude improvement in the target range compared to the other two waveforms.

\begin{figure}
  \centering
  \includegraphics[width=0.9\linewidth]{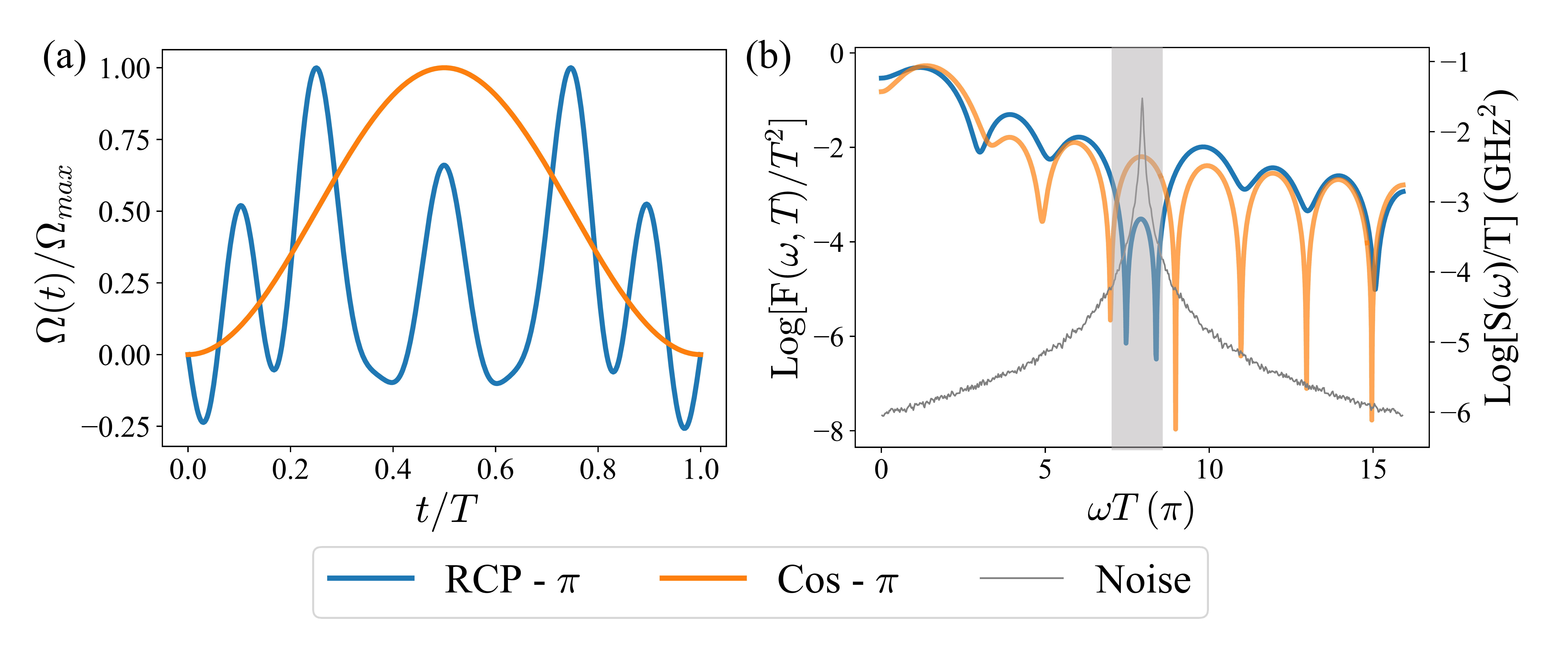}
  \caption{Optimized results of X gate for high-frequency noise. (a) The waveform of robust control pulse (RCP, blue) and Cos $\pi$-pulse (orange). (b) The filter function of RCP (blue) and Cos $\pi$-pulse (orange) overlapped with the corresponding noise spectrum (gray). }
  \label{FilterX_0.5G}
\end{figure}

Next is the optimization for high-frequency noise. We need to suppress an appropriate high frequency band $(\omega_1, \omega_2)$. Fig.\ref{FilterX_0.5G} shows the optimized results for $(3.6\omega_0,4.4\omega_0)$. The optimized waveform demonstrates at least an order of magnitude improvement over the Cos waveform in the target frequency band, with the angle of rotation still $\pi$.

Finally, in the case of mixed noise, multiple optimization intervals are selected according to the noise power spectrum distribution. As a demonstration, we choose the bands $(0, \omega_0)$ and $(2.5\omega_0,3.5 \omega_0)$.  Fig.\ref{FilterX_0+03}(c)(d) presents the optimization results. The optimized result also gives a $9\pi$ rotation. Here we compare it with Cos $9\pi$-pulse, Cos $\pi$-pulse, and static RCP. The suppression effect of the optimized waveform in the target range is more than an order of magnitude better than other waveforms.

Then we perform the robustness verification of these waveforms. We employ the inverse Fourier transform method, see Sec.\ref{App_NoiseModel}, to simulate a generalized Lorentz noise spectrum:
\begin{equation}\label{Lorentz_noise}
\frac {A^2}{\gamma+\frac{(\omega-\omega_0)^2}\gamma}
\end{equation}
Here, $\omega_0$ adjusts the peak position of the noise spectrum, and $\gamma$ controls the peak width of the spectrum. For different optimized waveforms, we choose distinct peak positions for the noise spectrum. To demonstrate the role of bandwidth, we simulated noise spectra of varying widths at each peak by adjusting $\gamma$ from 0.001 to 0.05. For each average fidelity data point, we averaged over 300 different random time-dependent noise signals. Fig.\ref{RobustnessCZ} (d) shows the width variation of the noise spectrum. It is worth noting that narrower noise spectra have higher peak values. However, to clearly display the width variations, the maximum value of each spectrum is normalized to one. All pulses selected for testing have a duration of 50 ns, corresponding to $\omega_0 = 2\pi/50$ GHz.

\begin{figure}
  \centering
  \includegraphics[width=0.9\linewidth]{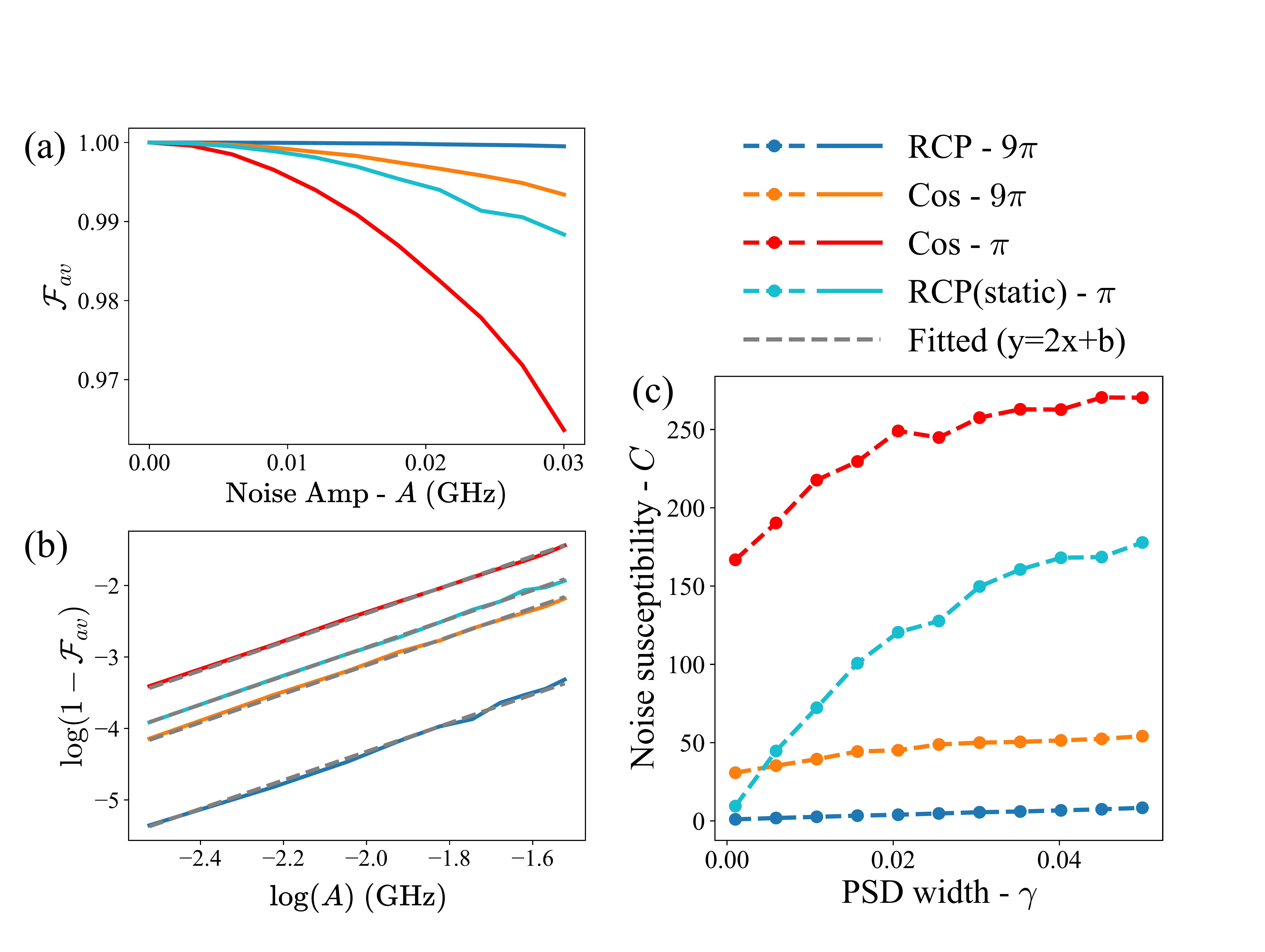}
  \caption{Robustness verification results of X gate for low-frequency noise. (a), (b) Normal and log version of average X-gate fidelity implemented by low-frequency RCP, Cos $\pi$-pulse, Cos $9\pi$-pulse, and static RCP against time-dependent noise with different noise amplitude.  (c) Noise susceptibility C against PSD width $\gamma$. Each data point is obtained by fitting average infidelity against noise amplitude as in (b).  And it is normalized by the minimized value of RCP pulse. }
  \label{RobustnessX_0}
\end{figure}

Fig.\ref{RobustnessX_0} shows the robustness verification results under low-frequency noise. We compare the low-frequency robust control pulse with the Cos $\pi$-pulse, the Cos 9$\pi$-pulse, and the static RCP. The noise amplitude $A$ is varied from 0 to 0.03$\Phi_0$. The peak position $\omega_0=0$. Fig.\ref{RobustnessX_0} (a),(b) plots the average gate fidelity versus noise amplitudes. The log graph satisfies $2x+b$ form very well, which here is $ \log( 1- \mathcal{F}_{avg}) \approx  2\log(A) + \log( C)$. Fig.\ref{RobustnessX_0} (c) displays the fitted noise susceptibility for different noise spectra. This graph shows that the robust pulse is much better than the other pulses across different PSD widths, due to the big bandwidth of the filter function. The static noise robust control pulse quickly loses its robustness when the noise spectrum becomes wider.

\begin{figure}
  \centering
  \includegraphics[width=0.9\linewidth]{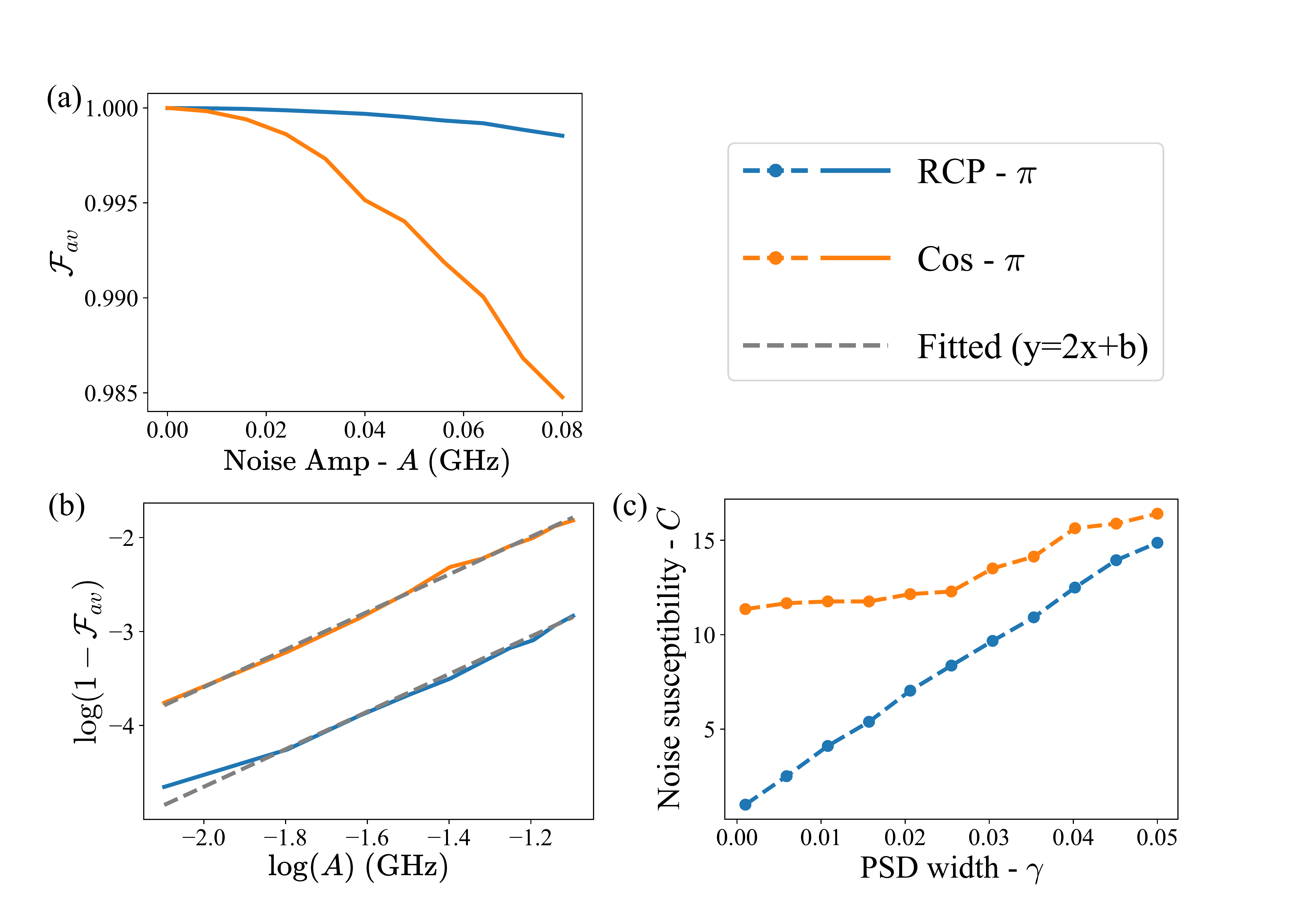}
  \caption{Robustness verification results of X gate for high-frequency noise. (a), (b) Normal and log version of average X-gate fidelity implemented by high-frequency RCP and Cos $\pi$-pulse against time-dependent noise with different amplitude.  (c) Noise susceptibility C against PSD width $\gamma$. Each data point is obtained by fitting average infidelity against noise amplitude as in (b). And it is normalized by the minimized value of RCP pulse.}
  \label{RobustnessX_0.5G}
\end{figure}

Fig.\ref{RobustnessX_0.5G} presents the robustness verification results under high frequency noise,  we compare it with the Cos $\pi$-pulse. Here the noise amplitude $A$ is varied from 0 to 0.08$\Phi_0$. A larger amplitude range is used here because the filter function of the Cos waveform is also primarily concentrated at low frequencies, giving it some suppression effect on high-frequency noise. The robust control pulse has a robustness advantage under different noise spectra. However, its noise susceptibility C grows rapidly with increasing PSD width due to its narrow filter function band.

\begin{figure}
  \centering
  \includegraphics[width=0.9\linewidth]{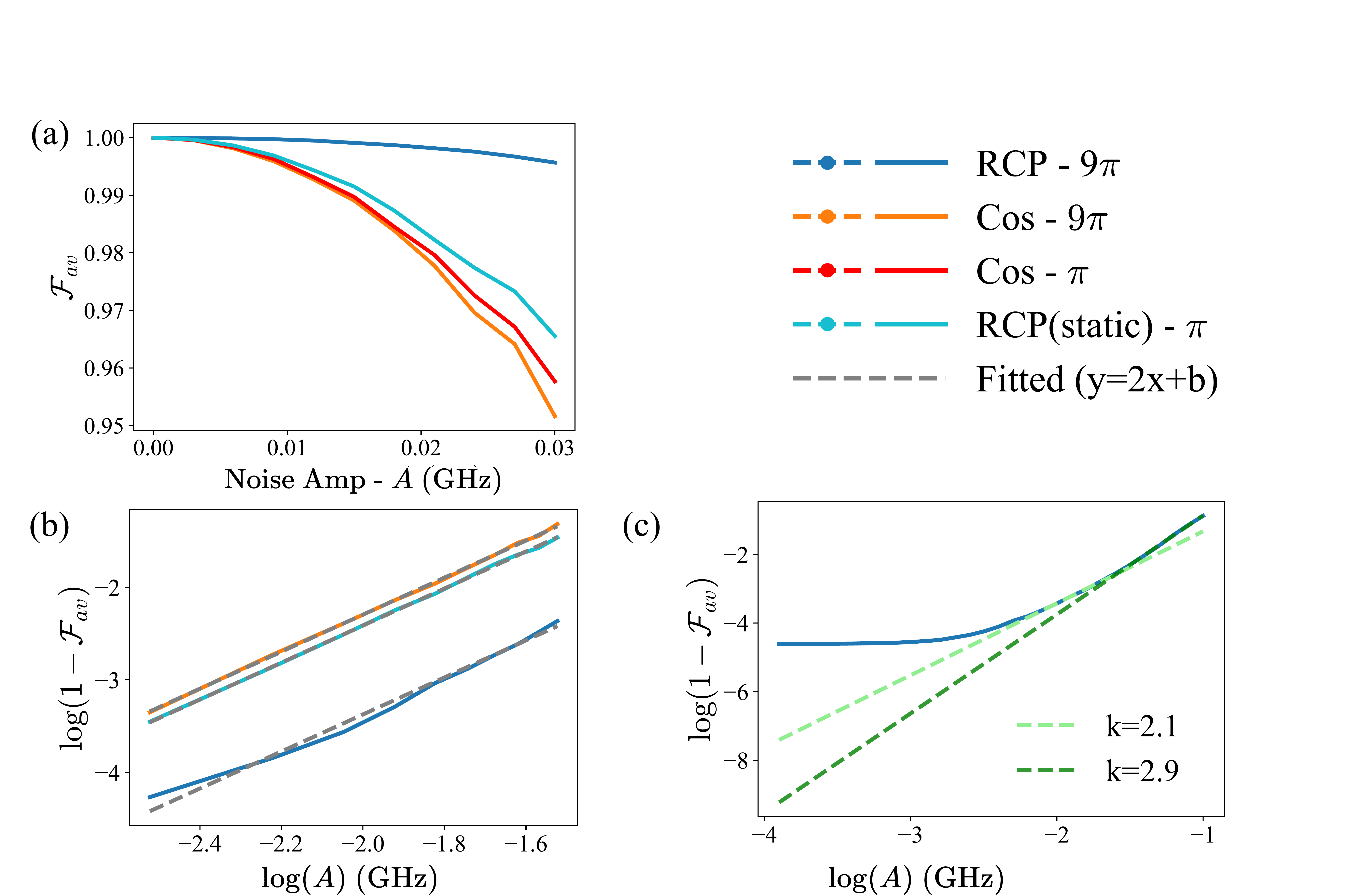}
  \caption{Robustness verification results of X gate for mixed noise. (a), (b) Normal and log version of average X-gate fidelity implemented by mixed noise RCP, Cos $\pi$-pulse, Cos $9\pi$-pulse, and static RCP against time-dependent noise with different amplitude.  (c) log infidelity of mixed noise RCP with a wider noise amplitude range.}
  \label{RobustnessX_03}
\end{figure}

Fig.\ref{RobustnessX_03} presents the outcomes of the robustness verification under mixed noise. Here we compared the mixed noise robust control pulse with the Cos $\pi$-pulse, the Cos 9$\pi$-pulse, and the static RCP. The noise amplitude is varied from 0 to 0.03$\Phi_0$. Mixed noise RCP has a clear advantage and static RCP cannot handle mixed noise scenarios. In the log graph of infidelity, we don't plot the curve of the Cos $\pi$-pulse, because it is between two very close curves. The infidelity of the mixed noise RCP does not match the fitting curve perfectly.  To explain why we simulate a wider noise amplitude range in Fig.\ref{RobustnessX_03}(c). 

For smaller noise amplitudes, the infidelity saturates at the noise-free limit; while for larger amplitudes, higher-order contributions become significant:
\[
\mathcal{F}_{av}\;=\;1 - b_2\,A^2 - b_4\,A^4 + O(A^6).
\]
Because we are dealing with Gaussian stationary noise, all odd moments of noise are zero. When $b_2$ is extremely small, the $b_4\,A^4$ term contributes more significantly, raising the slope of the infidelity curve.  This explains the deviation of the RCP infidelity from the simple linear fit with slope.

\section{Robust Two-Qubit Gate}\label{Sec_Rob2QBG}
As another practical application, we show a robust control-Z gate for fluxonium qubits~\cite{ma2024native} that efficiently dynamically corrects errors subject to flux noise by modulating the two-qubit coupling. Previous studies~\cite{hai2022universal,ma2024native} have identified this as a challenging task. 

The Hamiltonian of the inductively coupled fluxoniums
\begin{equation}
\begin{aligned}
H(\Phi_A,\Phi_B)=&\sum_{j=A,B}\frac{\omega^{j}(\Phi_j)}{2}\hat{Z}_j+J^{//}_{xx}\hat{X}_A\hat{X}_B \\ &+J^{//}_{zz}\hat{Z}_A\hat{Z}_B-\sum_{j\neq k}J^\times_{jk}\hat{Z}_j\hat{X}_k,
\end{aligned}
\end{equation}
where $\omega_j(\Phi_j)  = \sqrt{\epsilon_j^2+ \Delta_j^2}$, $J^\times_{jk}=J\frac{\epsilon_j\Delta_k}{\omega_A\omega_B}$,  $J^{//}_{xx}=J\frac{\Delta_A\Delta_B}{\omega_A\omega_B}$, $J^{//}_{zz}=J\frac{\epsilon_A\epsilon_B}{\omega_A\omega_B}$, and  $\epsilon_j(\Phi_j) =2I^j_p \Phi_j$. $I^j_p$ are stable system parameters in each fluxonium and represent the persistent superconducting loop current depending on their Josephson tunneling energies $\Delta_j$, also stable system parameters. Magnetic flux bias $\Phi_j$ on each of the fluxoniums is controlled by external fields, and hence is involved with flux noise $\delta \Phi_j$, which can be seen above couple to all the terms in the Hamiltonian. Without loss of generality, we assume the noise is small compared to the control field. The dependence of $\omega_j$ and $J^{\sim}$, $\sim\in\{//,\times\}$ on $\delta\Phi$ is:
\begin{equation*}
\begin{aligned}
&\delta \omega_j =2\frac{\epsilon_j}{\omega_j}I_p^j\delta\Phi_j+O(\delta\Phi_j^2)\\
&\delta J_{jk}^{\times} = 2J \frac{\Delta_k\Delta_j^2}{\omega_k\omega_j^3} I_p^j\delta\Phi_j - 2J \frac{\epsilon_j\epsilon_k\Delta_k}{\omega_j\omega_k^3} I_p^k\delta\Phi_k +O(\delta\Phi_j^2)\\
&\delta J_{xx}^{//} = \sum_{j\neq k} 2J \frac{\Delta_k\Delta_j\epsilon_j}{\omega_k\omega_j^3} I_p^j\delta\Phi_j+O(\delta\Phi_j^2)\\
&\delta J_{zz}^{//} = \sum_{j\neq k} 2J\frac{\epsilon_k\Delta_j^2}{\omega_k\omega_j^3} \frac{}{} I_p^j\delta\Phi_j+O(\delta\Phi_j^2)
\end{aligned}
\end{equation*}
Here $\delta\Phi_j$ is the original noise amplitude $\beta_j(t)$,  $\frac{\partial\omega_j}{\partial\Phi_j}\bigg|_{\Phi_j},\frac{\partial J^\sim}{\partial\Phi_j}\bigg|_{\Phi_j}$ are the noise coupling strength $\{c_{ji}(t)\}$ for the noise coupling operator of $\beta_j(t)$. 

The CZ gate in this system is realized by simultaneously detuning qubit A and B's frequencies. To simplify the calculation, we assume that $2I_p^{A,B}/\Delta_{A,B} =K$, $\Phi_{A,B}(t)=\Phi(t)$. Under the rotating frame of qubits' idle frequency $\omega_{0(A/B)}$, the elements in the xx term get high-frequency coefficients $e^{\pm i(\omega_B \pm \omega_A)t}$ and the elements in the zx term get high-frequency coefficients $e^{\pm i\omega_{A,B}t}$. When $\omega_{A,B},|\omega_A-\omega_B |\gg J$, we can take the RWA: 
\begin{equation}
H_0\approx\frac{1}{2}\Omega(t)\hat{Z}_A+\frac{1}{2}\eta\Omega(\Phi)\hat{Z}_B+J^{//}_{zz}(\Phi)\hat{Z}_A\hat{Z}_B
\end{equation}
Here, $\Omega(\Phi)=\Delta_j(\sqrt{K^2\Phi^2+1}-1)$ represents the qubit frequency detuning from the idle point.  The coefficient  $\eta=\frac{\Delta_B}{\Delta_A}$ is a constant. 

During this process, the qubit frequency shifts away from the sweet spot, making the qubit highly sensitive to flux noise~\cite{ma2024native}, Here we suppose these two qubits couple to same magnetic field environment, $\beta_j(t)=\beta(t)$, and the noise Hamiltonian takes the form:
\begin{equation}\label{2q_noise}
H_n=\beta(t)[\frac12c_{z_A}(t)\hat{Z}_A+\frac12c_{z_B}(t)\hat{Z}_B+c_{zz}(t)\hat{Z}_A\hat{Z}_B]
\end{equation}

Where $c_{z_j}(t)=\Delta_jK^2\Phi(t)/\sqrt{1+K^2\Phi^2(t)},c_{zz}(t)=2JK^2\Phi(t)/(1+K^2\Phi^2(t))^2$.  Under the interaction frame of $H_0$, $\widetilde{H}_n=H_n$. The filter function does not change with the control operation in the previous formalism. But here, we can transform $c(t)$ as the filter function. Since there is only one noise source, the approximate average fidelity takes the form :
\begin{equation}\label{approx_fid_S_CZ}
\mathcal{F}_{avg} \approx 1- \int \frac{d\omega }{2\pi} F(\omega,T)S(\omega)
\end{equation}
with the filter function $F(\omega,T)=\sum_{i}\left|\int_0^Tc_{i}(t)e^{-i\omega t}dt\right|^2, i\in\{z_A,z_B,zz\}$. The $c_{z_A},c_{z_B}$ parts of the filter function are used to suppress the noise impact on dephasing, while the $c_{zz}$ part is used to suppress the noise impact on the CZ phase.  The noise PSD $S(\omega)=\frac14\int \langle\beta(0)\beta(\tau)\rangle e^{-i\omega \tau}d\tau$, $\tau = |t_A-t_B|$. Where the coefficient 1/4 is from Eq.\ref{2q_noise}. 
\begin{figure}
  \centering
  \includegraphics[width=0.6\linewidth]{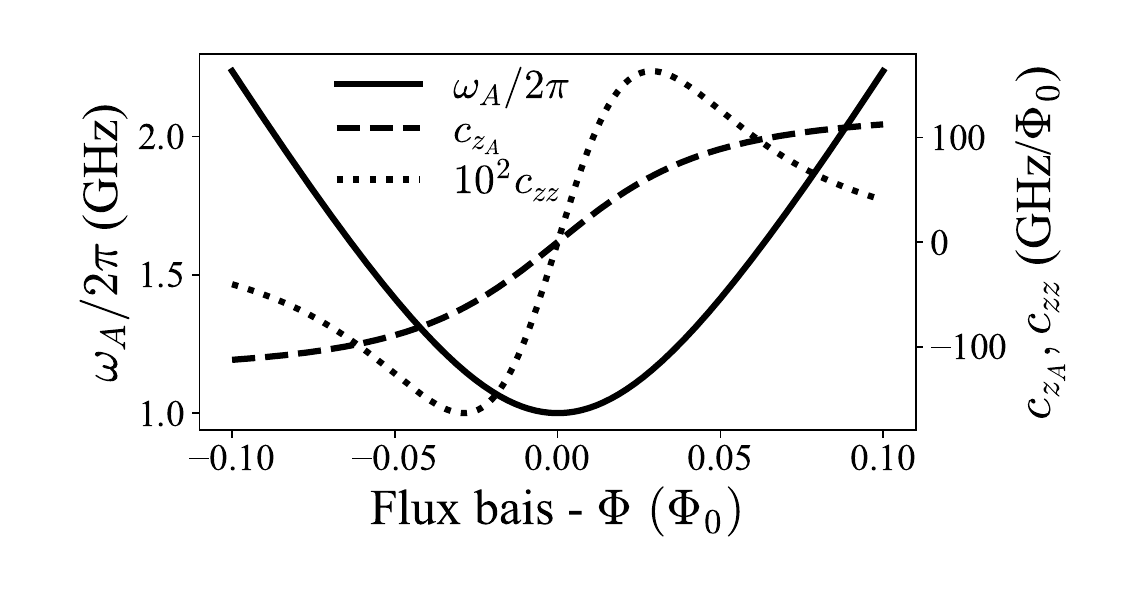}
  \caption{The change of Qubit A's frequency and noise coupling strength $c_{z_A},c_{zz}$ to magnetic flux.} 
  \label{Omega_phi}
\end{figure}

\begin{figure}
  \centering
  \includegraphics[width=0.8\linewidth]{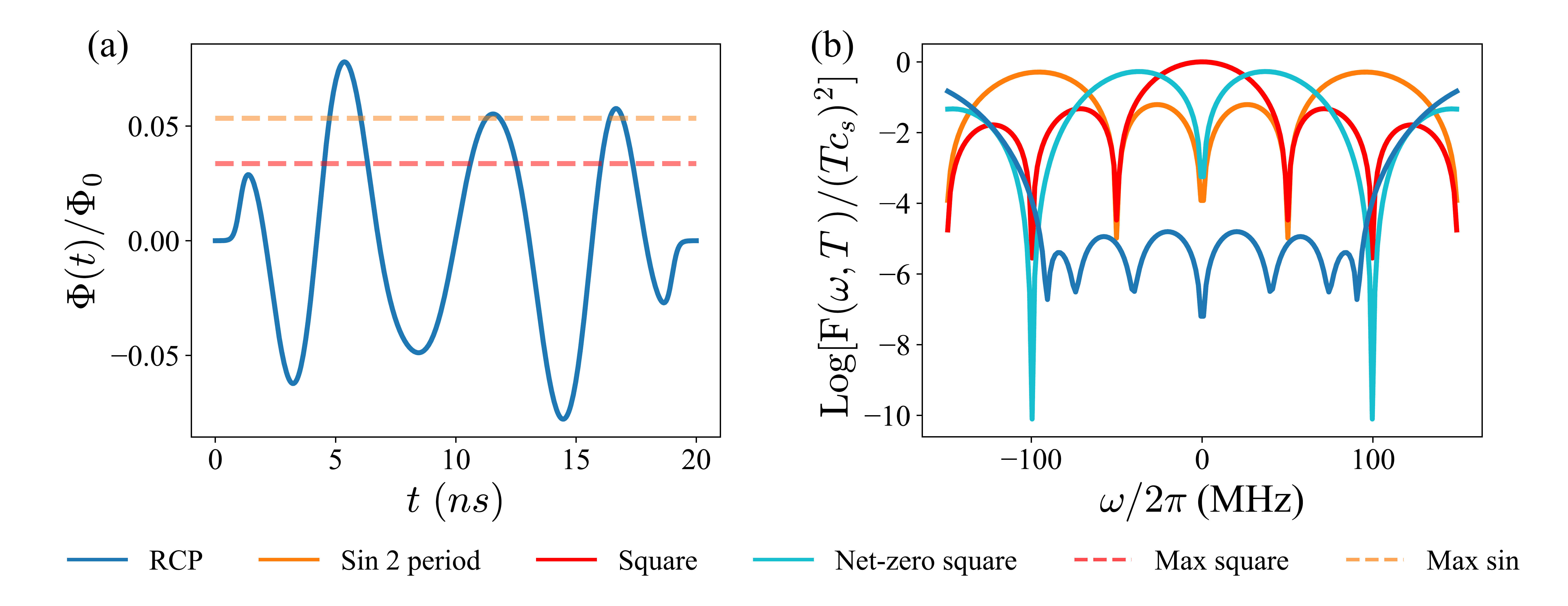}
  \caption{Optimized results of CZ gate. (a) The pulse waveform of robust control pulse(RCP) and dotted lines are the maximum amplitude of sine and square pulse. (b)The log diagram of the filter function of RCP and other commonly used pulses. } 
  \label{FilterCZ}
\end{figure}

In this article, we design a 20ns CZ pulse with a target filter function band $(0,100)\rm MHz$. The parameters  are $\Delta_{A(B)}=1(1.3)\mathrm{GHz}$, $\epsilon_{A(B)}=\frac{20*\Delta_{A(B)}}{\Phi_0}\Phi$ here. The corresponding frequency spectrum and noise coupling strength are shown in Fig.\ref{Omega_phi}. The optimized waveform and filter function are presented in Fig.\ref{FilterCZ}. Compared to other waveforms commonly used in experiments, we find that its filter effect is at least two orders of magnitude better than other waveforms, while the amplitude and complexity of the optimized waveform remain at the same level as commonly used waveforms. 

\begin{figure}
  \centering
  \includegraphics[width=0.9\linewidth]{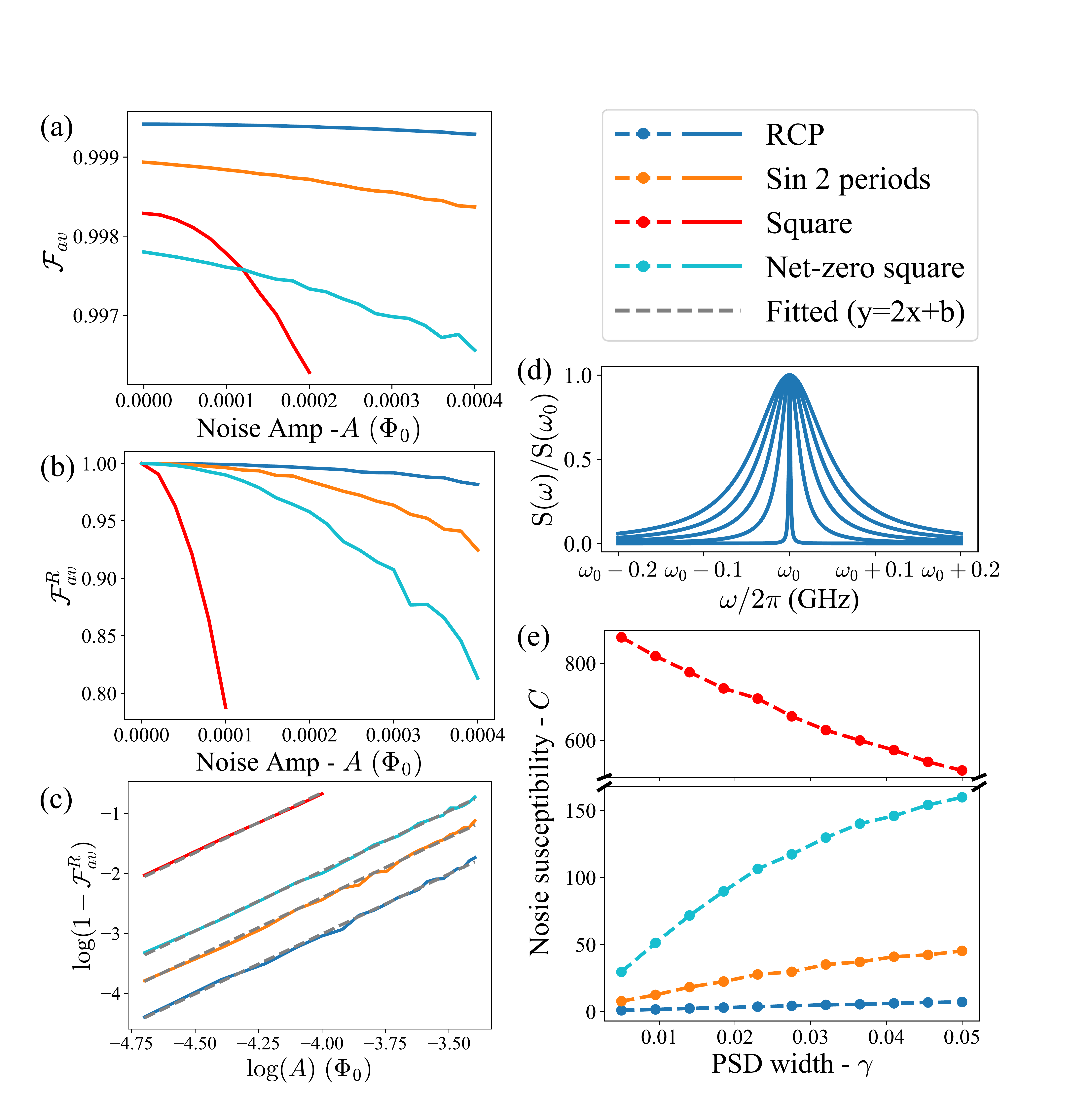}
  \caption{ Robustness verification results of CZ gate. (a) Average CZ gate fidelity implemented by RCP, Cos $\pi$-pulse and Cos $9\pi$-pulse against time-dependent noise with different noise amplitude. (b)(c) Normal and log version of robustness(self-compared average gate fidelity) against different noise amplitude. (d) Noise susceptibility against PSD width $\gamma$. Each data point is obtained by fitting average infidelity against noise amplitude as in (c).  And it is normalized by the minimum value of RCP pulse}
  \label{RobustnessCZ}
\end{figure}

We show the robustness verification results of these waveforms in Fig.\ref{RobustnessCZ}. Noise in (a), (b), and (c) is $1/f$ noise generated by an ensemble of RTN processes, see Sec.\ref{App_NoiseModel}. For each data point, we average over 300 different time-dependent noise signals. One important point to note in Fig.\ref{RobustnessCZ}(b) is that the advantage of the robust waveform is not only the robustness but also the noise-free gate fidelity. (noise-free gate fidelity is the gate fidelity achieved only by the control waveform when noise is absent.) Due to the coupling terms other than the $ZZ$ term, constructing an ideal CZ gate is challenging. The commonly used waveforms only have one freedom, 
 the overall amplitude, to be optimized, whereas the robust waveform allows more freedom in pulse shaping. If the shape of the waveform is fixed that we only optimize the amplitude, there is an upper limit to the fidelity. To eliminate the effects of the noiseless gate fidelity, we extract pure robustness in Fig.\ref{RobustnessCZ}(b)(c) by calculating the self-compared gate fidelity, $\mathcal{F}^R_{av}$, which substitutes $U_G$ by $U_0$ in Eq. \ref{fid_av}. Where the robust waveform demonstrates a clear advantage. 

We obtain the noise susceptibility by fitting the log infidelity in Fig.\ref{RobustnessCZ}(c).  Then we simulated Lorentz-type noise picked at zero with varying widths to demonstrate the filter function's wider suppression bandwidth, as shown in Fig.\ref{RobustnessCZ}(e). We observe that the noise susceptibility of the robust pulse does not increase significantly with the noise PSD width, whereas it shows substantial growth for the Sin two-period and Net-zero square pulses. Notably, the square pulse is particularly sensitive to low-frequency noise, with higher-frequency noise tending to cancel itself out. This leads to the noise susceptibility of the square pulse decreasing as the PSD widens, which is also evident from the filter function spectrum.

\section{Enhanced Sensing of AC Signals}

\subsection{Single-Qubit Hamiltonian and Flux Control}

Consider a single-qubit sensor subjected to an AC signal. The system Hamiltonian is given by
\begin{equation}
H \;=\; \frac{\omega\bigl(\Phi(t)\bigr)}{2}\,\hat{Z},
\end{equation}
where $\Phi(t)$ denotes the external magnetic flux. This flux comprises (i)~a qubit bias $\Phi_b(t)$, (ii)~a flux noise term $\beta(t)$, and (iii)~an AC signal $\Phi_{s}(t) = B\,\cos\bigl(\omega_{s}\,t + \alpha\bigr)$.Here $B$ is the amplitude of the target signal to be measured. Assume $\Phi_b(t)\gg\Phi_s(t),\beta(t)$, the qubit frequency can be approximated as: $\omega(\Phi(t))\approx \omega(\Phi_b(t))+\delta\omega_n(t)+\delta\omega_s(t)$, where
\begin{equation}
\begin{aligned}
    &\delta\omega_n(t) = c_{z}(t)\,\beta(t)\\
    &\delta\omega_s(t) = B\,c_{z}(t)\,\cos\bigl(\omega_{s}\,t + \alpha\bigr).
\end{aligned}
\end{equation}
$c_{z}(t)$ quantifies the flux-to-qubit energy shift
\begin{equation}
c_{z}(t) 
\;=\;
\frac{4\,\Delta\,I_{p}^{2}\,\Phi(t)}{\sqrt{\Delta^{2} \;+\; 4\,I_{p}^{2}\,\Phi^{2}(t)}},
\end{equation}
where $\Delta$ and $I_{p}$ are hardware-dependent energy and current parameters, respectively.

\subsection{Estimate method and Fisher Information}

\paragraph{Initial State and Measurement Outcomes.}
We prepare the qubit in the state
\[
\lvert + \rangle \;=\; \frac{1}{\sqrt{2}}\bigl(\lvert 0 \rangle + \lvert 1 \rangle\bigr),
\]
and measure in the $\{\lvert + \rangle,\lvert - \rangle\}$ basis. In the presence of the AC signal of amplitude $B$, the measurement probabilities are
\begin{equation}
\begin{aligned}
P_{+}(B) &= \cos^{2}\!\Bigl(\tfrac{B\,\Gamma(t)}{2}\Bigr), \\
P_{-}(B) &= \sin^{2}\!\Bigl(\tfrac{B\,\Gamma(t)}{2}\Bigr).
\end{aligned}
\end{equation}
Here,
\begin{equation}
\Gamma(t)
\;=\;
\int_{0}^{t} c_{z}(v)\,\cos\bigl(\omega_{s}\,v+\alpha\bigr)\,dv
\end{equation}
characterizes how the control function $c_{z}(t)$ and the target frequency $\omega_{s}$ together modulate the phase accumulation of the qubit during the sensing interval $[0,t]$. This phase doesn't include the contribution of reference frequency $\omega(\Phi_b(t))$. One can estimate B from the expected value $\langle \sigma_{x}\rangle \;=\; \cos\bigl(B\,\Gamma(t)\bigr),$. 

Most sensing processes are performed under multicolored noise. To reduce the error caused by noise, in the Gaussian stationary noise condition, we can rederive the measurement probabilities as:
\begin{equation}
\begin{aligned}
    \langle P_+(B)\rangle=& \bigl\langle \cos^{2}\!\Bigl(\tfrac{B\,\Gamma(t)+ \phi_n}{2}\Bigr)\bigr\rangle\\
    =&\bigl \langle \frac14(2+e^{i(B\,\Gamma(t)+ \phi_n)}+e^{-i(B\,\Gamma(t)+ \phi_n)})\bigr\rangle\\
    =&\frac14\bigl(2+ e^{\langle  \phi_n^2 \rangle/2} ( e^{iB\,\Gamma(t)}+e^{-iB\,\Gamma(t)})\bigr)\\
    =&\frac12(1+e^{\langle  \phi_n^2 \rangle/2}\cos(B\,\Gamma(t))\\
    \\
    \langle P_-(B)\rangle=&1-\langle P_+(B)\rangle\\
    =&\frac12(1-e^{\langle  \phi_n^2 \rangle/2}\cos(B\,\Gamma(t))
\end{aligned}
\end{equation}
where $\langle\cdot\rangle$ denote the average for environment noise, $\phi_n=\int_0^t c_z(v)\beta(v)dv$ is the phase term accumulated by noise.  We use the relation, $\langle  e^{i\phi_n} \rangle=e^{\langle  \phi_n^2 \rangle/2}$ in our derivation, since $\phi_n$ is a Gaussian random variable. 

\paragraph{Signal Estimation and Fisher Information.}
The estimation scheme here is adjusted from the scheme for the dynamic decoupling sequence in \cite{degen2017quantum}.
We let the phase accumulated by the signal satisfy
\begin{equation}
    \phi_s(t) = B\,\Gamma(t)=B\bar{\Gamma}t,\ \  \bar\Gamma=\Gamma(T)/T
\end{equation}
at times $t=nT$. In that case, we can get B from the estimation of the Ramsey oscillation frequency, $\omega_{est}=B\bar{\Gamma}$, of $\langle P_+(B)\rangle$. The calculation is similar if there are signals of multiple frequencies. 
\begin{equation}
   \bar \Gamma=\sum\bar{\Gamma}_i t=\sum_i\Gamma_i(T)/T
\end{equation}
where $\Gamma_i(T)\;=\;\int_{0}^{T} c_{z}(v)\,\cos\bigl(\omega_{i}\,v+\alpha_i\bigr)\,dv$. This scheme yields that $c_z(v)\cos (\omega_i v+\alpha)$ has a period $T$.
\begin{equation}
    c_{z}(v)=c_{z}(v+T), \omega_iT=2k\pi
\end{equation}
This means a control waveform with periodic repetition length T can potentially measure  the signals located in a frequency comb $\{\omega_k=k\frac{2\pi}{T}\}$.  And the period T detemines the resolution ratio of the detection.
Every frequency $\omega_i$ should satisfy the periodic condition. In practice, we can adjust the size of $T$ according to the frequency to be measured.

The Fisher information with respect to $B$ can be shown as:
\begin{equation}
\begin{aligned}
\mathcal{I}(B) &=\sum_n \frac{e^{-\langle\phi_n^2\rangle} \sin^2(B\bar{\Gamma}nT)}{1 - e^{-\langle\phi_n^2\rangle}+e^{-\langle\phi_n^2\rangle} \sin^2(B\bar{\Gamma}nT)}(\bar{\Gamma}nT)^2\\
&=\sum_n \frac{1}{1 + \frac{1/e^{-\langle\phi_n^2\rangle} - 1}{ \sin^2(B\bar{\Gamma}nT)}}(\bar{\Gamma}nT)^2
\end{aligned}
\end{equation}
Since the value of $B$ is not known a priori, the period term $\sin^{2}(B\bar{\Gamma}nT)$ cannot be optimized. Thus, we enhance sensitivity to $B$ by enlarging the value of $\bar{\Gamma}$ and decreasing the value of $\langle \phi_n^2\rangle$.

\subsection{Frequency Perspective and Filter Function}

\paragraph{Maximizing the Sensing Response.}
One can express $\Gamma(t)$ using a Fourier transform. 
\begin{equation}
\begin{aligned}
     \Gamma(t)&=\int_0^t c_z(v)\cos (\omega_s v+\alpha)dv\\&=1/2\int_{-\infty}^{\infty}[e^{-i\alpha}\delta(\omega+\omega_s)+e^{i\alpha}\delta(\omega-\omega_s)]c_z(-\omega,t)d\omega\\&=\cos(\theta_s-\alpha) |c_z(\omega_s,t)| . 
\end{aligned}
\end{equation}
If the signal phase $\alpha$ can be freely chosen, one can set $\alpha = \theta_{s}$ (the phase angle of $c_{z}(\omega_{s}, t)$) to ensure $\Gamma(t)$ is real and maximized:
\[
\Gamma(t)
\;=\;
\bigl\lvert c_{z}(\omega_{s}, t)\bigr\rvert,
\quad
c_{z}(\omega, t)
\;=\;
\int_{0}^{t} 
c_{z}(v)\,
e^{-\,i\,\omega\,v}\,dv.
\]
Thus, for optimal frequency-selective sensing, one aims to maximize $\int |c_{z}(\omega,t)|^{2} d\omega$ within the target frequency range.

\paragraph{Noise and Filter Function.}
The influence of noise $\langle\phi_n^2\rangle$ can be expressed as:
\begin{equation}
\begin{aligned}
\langle\phi_n^2(t)\rangle
&=\;
\int_{0}^{t}
\int_{0}^{t}
\langle \beta(t_{1})\,\beta(t_{2})\rangle \,
c_{z}(t_{1})\,c_{z}(t_{2})
\,dt_{1}\,dt_{2}\\
&=\;
\int_{-\infty}^{\infty}
S(\omega)\,F(\omega,t)\,d\omega,
\end{aligned}
\end{equation}
and $S(\omega)$ is the noise power spectral density. The filter function is
\begin{equation}
F(\omega,t)
\;=\;
\bigl\lvert c_{z}(\omega,t)\bigr\rvert^{2}
\;=\;
\biggl\lvert
\int_{0}^{t}
c_{z}(v)\,e^{-\,i\,\omega\,v}\,dv
\biggr\rvert^{2}.
\end{equation}
To reduce decoherence, one must suppress $F(\omega,t)$ in noise frequency regions. 

\paragraph{Optimization Strategy.}
Therefore, the control design $c_{z}(t)$ must achieve:
\begin{equation}
\max
\sum_{\mathcal{B}_{i}}
\int_{\mathcal{B}_{i}}
F(\omega,t)\,d\omega
\quad\text{and}\quad
\min
\sum_{\mathcal{A}_{i}}
\int_{\mathcal{A}_{i}}
F(\omega,t)\,d\omega,
\end{equation}
where $\mathcal{B}_i$ labels the target signal bands and $\mathcal{A}_i$ labels the noise-dominated bands.

\subsection{High accuracy correction}
The approximation that $\omega(\Phi_b+\Phi_{n/s})\approx c_z(t)\Phi_{n/s}$, when $\Phi_{n/s} \ll \Phi_b$, is enough for the above analysis. But if we want to further improve the sensing accuracy, we need to consider the corrections brought about by higher-order terms of $\Phi_{n/s}$. The higher-order terms of the signal will introduce additional phase accumulation. The higher-order terms of the noise will cause the mean frequency fluctuation at each moment to deviate from zero. This will also introduce additional phase accumulation.

The exact frequency change caused by signal and noise takes the form:
\begin{equation}
\delta\omega(t)=\omega(\Phi_b(t)+\Phi_s(t)+\Phi_n(t))-\omega(\Phi_b(t))
\end{equation}

The accurate phase accumulated by signal and noise, under the periodic conditions we set before, takes the form:
\begin{equation}
\phi(NT)=\int_0^{NT}\delta\omega(t)dt = \bar{\delta\omega}NT
\end{equation}
$\bar{\delta\omega}$ is the oscillation frequency to be estimated. 

In Gaussian stationary noise condition, we have $\Phi_n(t)  \sim N(0,\sigma)$, which is independent of t.  Then the statistic mean value of $\delta\omega(t)$ and $\bar{\delta\omega}$ can be expressed as:
\begin{equation}
\begin{aligned}
&\langle \delta\omega(t) \rangle = \int[\omega(\Phi_b(t)+\Phi_s(t)+\Phi_n)-\omega(\Phi_b(t)]P(\Phi_n)d\Phi_n\\
&\langle \bar{\delta\omega} \rangle = \frac{1}{T}\int_0^T\langle \delta\omega(t) \rangle dt
\end{aligned}
\end{equation}
After estimating the value of $\sigma$, we can solve the value of B from the estimated value of omega $\omega_{est}=\langle \bar{\delta\omega} \rangle$, since we already know $\Phi_s(t)$. The noise standard deviation $\sigma$ can be estimated similarly to the signal. In the absence of the signal, the frequency fluctuation becomes:
\begin{equation}
\langle \delta\omega(t) \rangle = \int[\omega(\Phi_b(t)+\Phi_n)-\omega(\Phi_b(t)]P(\Phi_n)d\Phi_n
\end{equation}
By estimating $\langle \bar{\delta\omega} \rangle$, we can solve the value of noise std $\sigma$. 

\subsection{Numerical Validation and Comparison with PDD}

\begin{figure}
  \centering
  \includegraphics[width=0.9\linewidth]{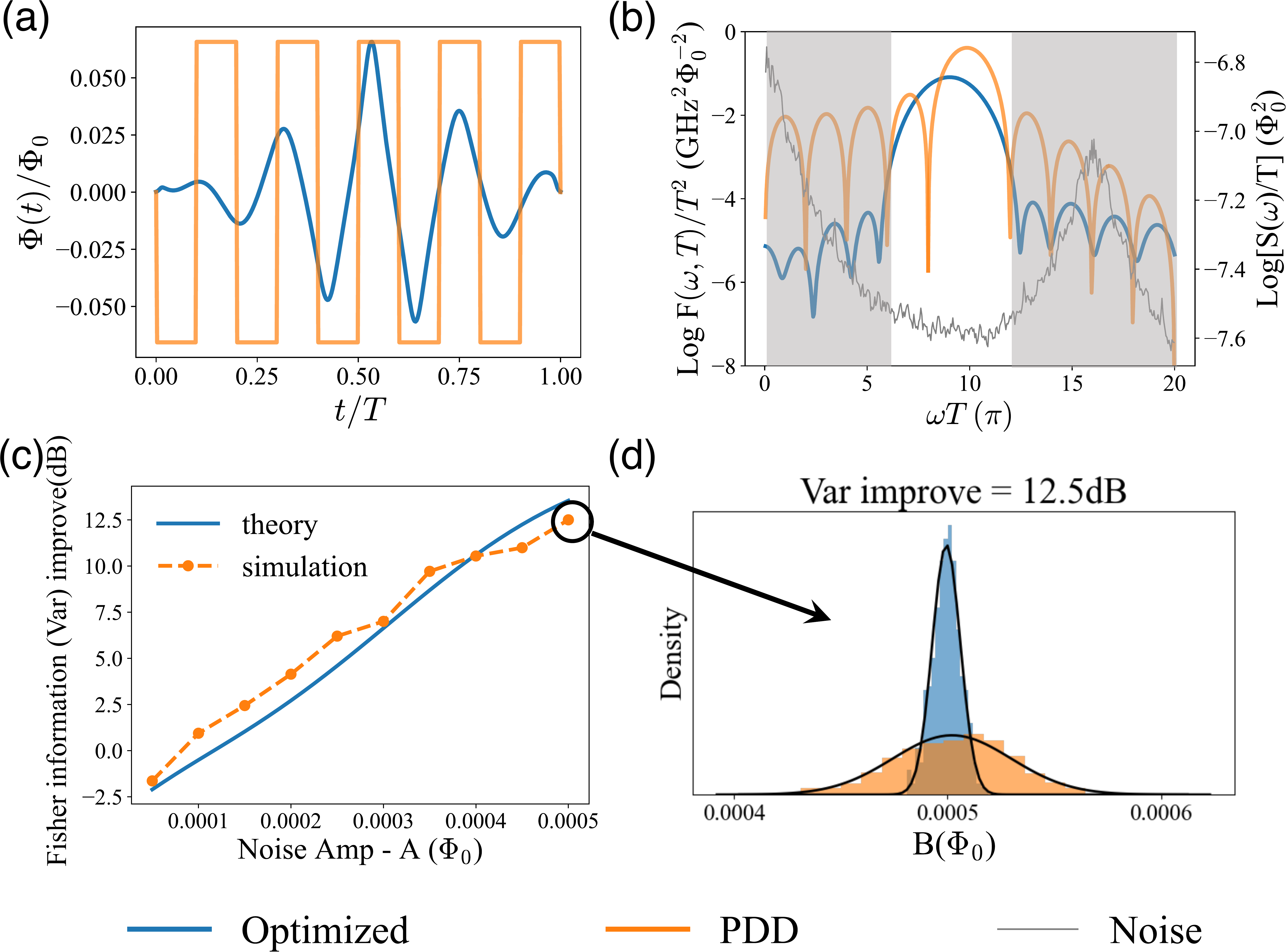}
  \caption{Results of the optimized sensing. (a) Flux bias $\Phi(t)$ for the optimized control and PDD. (b) The corresponding filter functions in a logarithmic scale, highlighting noise-suppression (shaded) and signal-amplification (unshaded) regions.  (c) the improvement in the variance of the simulation results of the optimized waveform versus PDD, as well as the improvement in the theoretical fisher information. (d) The Gaussian distribution of the measured $B$ values under each control scheme.}
  \label{Results_sensing_S}
\end{figure}
As a practical illustration, one can select unwanted noise ranges (e.g., $(0,3\omega_{0})\cup(6\omega_{0},10\omega_{0})$) and a desirable signal set $\{3\omega_0, 4\omega_0,5\omega_0,6\omega_0\}$ with a repeat measurement period $T=\frac{2\pi}{\omega_0}$.
Fig.~\ref{Results_sensing_S}(a) shows the flux bias of the optimized flux-control strategy compared to the periodic dynamical decoupling (PDD) sequence with a chosen pulse interval $\tau$. This PDD sequence can amplify signals at frequency $\frac{1}{(2\tau)}$. We set the frequency of PDD equal to one of the target signal frequencies. The displayed example here is $\frac{2\pi}{(2\tau)}=5\omega_0$. By replacing $\pi$-pulses with a rapid flux switching (a square waveform for $c_{z}(t)$), one can mimic the PDD filter function without the need for fast gate operations. Fig.~\ref{Results_sensing_S}(b) compares the optimized flux-control filter function to the PDD filter. 

\begin{figure}
  \centering
  \includegraphics[width=0.9\linewidth]{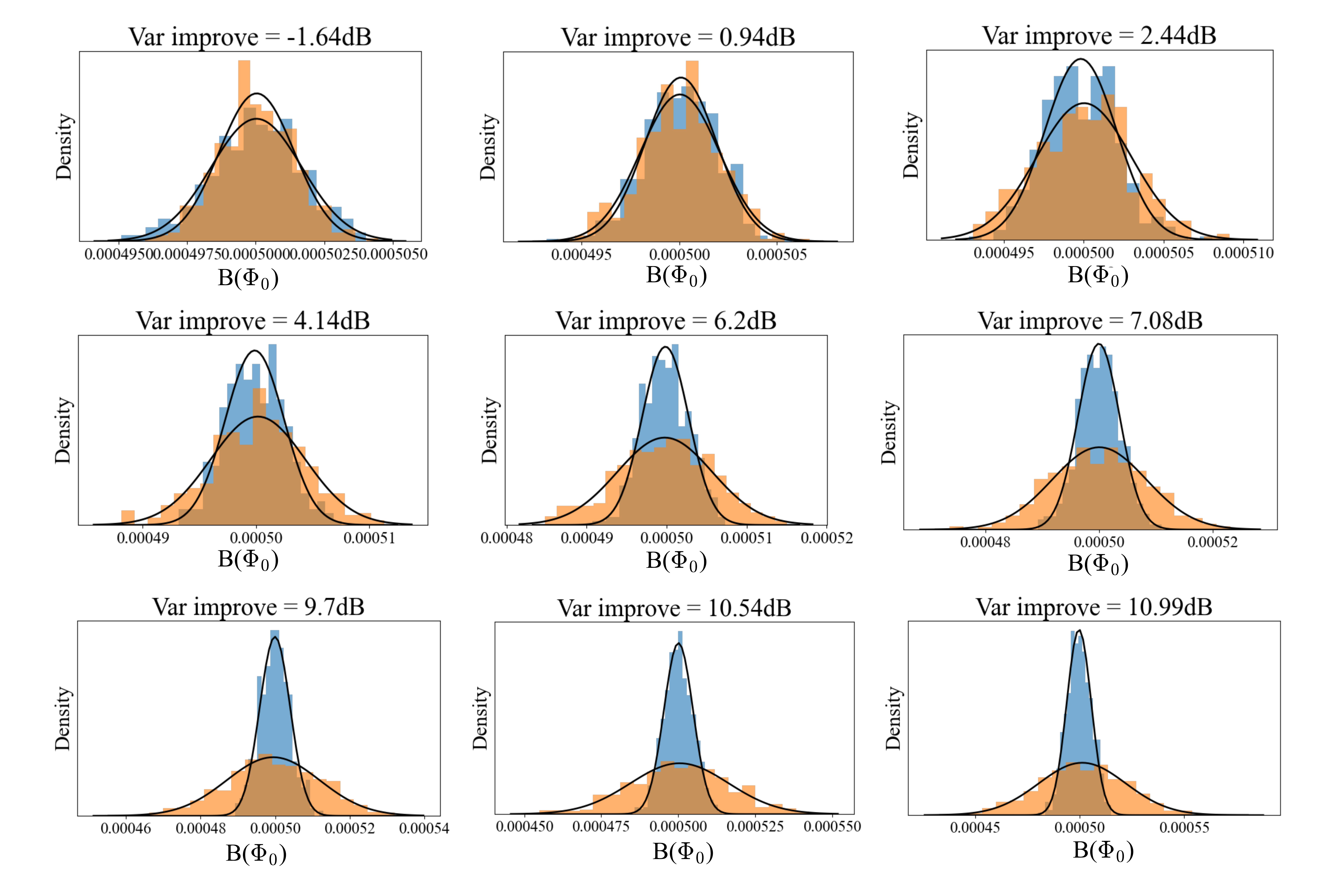}
  \caption{Statistical results of magnetic field strength estimation. From left to right and top to bottom, the noise amplitudes range from \(0.00005\Phi_0\) to \(0.00045\Phi_0\) in increments of \(0.00005\Phi_0\).}
  \label{Statistic_sensing}
\end{figure}

Fig.~\ref{Results_sensing_S}(c)(d) depicts the simulation results of the sensing process, where the qubit experiences Lorentzian noise peaks at $0$ and $8\,\omega_{0}$. In the simulation, the repeat period was chosen to be $T=100ns$. We take a single-shot measurement at times $T,2T,...,NT$, which gives the results of 0 or 1. We gather the outcomes of 300 single shots to make one estimate. Subsequently, we replicate this estimation process 600 times to obtain a statistical distribution. Fig.~\ref{Results_sensing_S}(c) demonstrates the improvement in the variance of the simulation results of the optimized waveform versus PDD as noise amplitude $A$ increases, as well as the improvement in the theoretical Fisher information. Numerical results confirm that the optimized flux-control strategy outperforms PDD as the noise amplitude gradually increases, achieving more than 10\,dB reduction in variance when noise amplitude is greater than $0.0004\Phi_0$. The estimated result of noise amplitude $A=0.0005\Phi_0$ is displayed in Fig.~\ref{Results_sensing_S}(d). The statistical results for other noise amplitudes are shown in Fig.~\ref{Statistic_sensing}.

In summary, to enhance AC signal sensing in a flux qubit, one tailors the flux-control waveform $c_{z}(t)$ such that it (i)~selectively amplifies signals in a chosen frequency band and (ii)~suppresses noise-induced dephasing elsewhere. This approach, analyzed via filter functions and Fisher information, leads to improved measurement accuracy and noise robustness over standard periodic dynamical decoupling methods.

\end{document}